\documentclass{aa}
\usepackage{mhchem}
\usepackage{natbib}
\usepackage{graphicx}
\usepackage{txfonts}
\usepackage{amsmath}
\usepackage{appendix}
\usepackage{tabularx}
\begin{document}

   \title{Chemical evolution in planet-forming regions with growing grains}

   \author{Christian Eistrup
          \inst{\ref{inst1}}\inst{\ref{inst2}}\thanks{Virginia Initiative on Cosmic Origins (VICO) Fellowship}
          \and
          L. Ilsedore Cleeves
          \inst{\ref{inst2}}
          \and
          Sebastiaan Krijt
          \inst{\ref{inst3}}
          }

   \institute{Max Planck Institute for Astronomy (MPIA), K\"onigstuhl 17, Heidelberg, Germany. \email{eistrup@proton.me}\label{inst1}
   \and
    Department of Astronomy, University of Virginia, 530 McCormick Road, Charlottesville, VA 22904, USA. \label{inst2}
    \and
    School of Physics and Astronomy, University of Exeter, Stocker Road, Exeter EX4 4QL, UK.\label{inst3}}

   \date{\today}

  \abstract
   {Planets and their atmospheres are built from gas and solid material in protoplanetary disks. This solid material grows from smaller, micron-sized grains to larger sizes in the disks, during the process of planet formation. This solid growth may influence the efficiency of chemical reactions that take place on the surfaces of the grains, and in turn affect the chemical evolution that the gas and solid material in the disk undergoes, with implications for the chemical composition of the planets.} 
   {Our goal is to model the compositional evolution of volatile ices on grains of different sizes, assuming both time-dependent grain growth and  several constant grain sizes. We also examine the dependency on the initial chemical composition.}
   {The state-of-the-art Walsh chemical kinetics code is utilised for modeling chemical evolution. This code has been upgraded to account for the time-evolving sizes of solids. Chemical evolution is modelled locally at four different radii in a protoplanetary disk midplane (with associated midplane temperatures of 120K, 57K, 25K and 19.5K) for up to 10Myr. The evolution is modelled for five different constant grain sizes, and one model where the grain size changes with time according to a grain growth model appropriate for the disk midplane.}
   {Local grain growth, with conservation of total grain mass and the assumption of spherical grains, acts to reduced the total grain-surface area that is available for ice-phase reactions. This reduces these reactions efficiency compared to a chemical scenario with a conventional grain-size choice of 0.1$\mu$m. The modelled chemical evolution with grain growth lead to increased abundances of \ce{H2O} ice. For carbon, in the inner disk grain growth leads CO gas to overtake \ce{CO2} ice as dominant carrier, and in the outer disk, \ce{CH4} ice to become the dominant carrier. Larger grain sizes leads to the C/O ratio in the gas phase changing less over time, than when considering 0.1$\mu$m-sized grains. Overall, a constant grain size adopted from a grain evolution model leads to almost identical chemical evolution, when compared to chemical evolution with evolving grain sizes. A constant grain size choice, albeit larger than 0.1$\mu$m, may therefore be an appropriate simplification when approximating the impact of grain evolution into chemical modelling.}  
   {}


   \maketitle
%


\section{Introduction} \label{sec:intro}
Planets form in protoplanetary disks around young stars. Exoplanet atmospheres can be observed, chemically characterised, and molecular constituents of the atmospheres can be identified. More generally, carbon-to-oxygen (C/O) ratios of giant exoplanet atmospheres can be retrieved from directly imaged exoplanets with existing facilities, such as the VLTI \textsc{Gravity}-experiment \citep[e.g., the exoplanets $\beta$ Pic b and HR 8799e, see][]{gravity2020betapic,molliere2020} and the \textsc{IGRINS}-instrument at the Gemini-South Observatory, which was used to retrieve the C/O ratio of the day-side hemisphere of the transiting exoplanet WASP-77Ab \citep[see][]{line2021ctoo}. Such chemical characterisation of exoplanet atmospheres will be accelerated by \emph{JWST}, both in terms of precision (lower uncertainty on C/O ratios) and in terms of the number of exoplanets for which atmospheres will be chemically characterised ($\sim$88)\footnote{Number of unique exoplanets targeted for atmospheric spectroscopic characterisation as described in Public PDFs of ERS, GTO, and Cycle 1 GO programs on Space Telescope Science Institute's \emph{JWST} portal (``Exoplanet''-category).}.

Observations of atmospheric compositions of exoplanets are a window into understanding what the atmospheres, and the exoplanets as a whole, formed from \citep[see e.g.][]{mordasini2016}. The idea is that the elemental chemical compositions of exoplanet atmospheres should reflect the combined elemental compositions of the material(s) that went into forming the exoplanet atmospheres. These are the materials that exist in protoplanetary disks, and which go into forming planets: gas and solids, where solids have both a refractory (rocky, with high desorption temperatures/molecular binding energies) and a volatile (icy, with low desorption temperatures/molecular binding energies) component. The region of protoplanetary disk midplanes where volatile species start to freeze out as ices usually have temperatures below 150K. \ce{H2O}, as the volatile molecule with the highest molecular binding energy \citep[see, e.g.,][]{fraser2001,penteado2017}, freezes out below this temperature.

Giant exoplanet atmospheres may be formed by run-away accretion of gas surrounding a solid planetary core \citep[as modelled by][with, respectively, without protoplanetary disk chemical evolution]{madhu2014,notsu_eistrup2020}, and by accretion of ice-covered solids, such as pebbles \citep[see, e.g.,][]{lambrechts2012,bitsch2019pebblesgiants,trapman2019}. Accretion of pebbles could lead to the material contained in the pebbles becoming part of the forming planet, and its atmospheres. In other words, the accretion of pebbles by exoplanets could affect the chemical makeup of the planetary atmospheres.

In order to establish a link between the chemical composition of a planetary atmosphere, and the formation history of the planet and its atmosphere, it is important to constrain the chemical composition of the volatile ices carried on these pebbles, as they drift from the outer, colder regions of the disk, inwards. The volatile ices are of special importance, since these ice components are the main carriers of carbon, oxygen, nitrogen and sulphur, which go into forming the gas-phase phase species constituting exoplanet atmospheres. Pebble drift in protoplanetary disks is an active area of research \citep[e.g.,][]{brauer2008,birnstiel2012,birnstiel2015}, including the accretion of pebbles onto planets \citep{mordasini2016,bitsch2015,bitsch2016,bitschtrifonov2020,bitschbattistini2020,cridland2016,cridland2017}, and how this accretion affects the resulting compositions of the exoplanet atmospheres \citep[see, e.g.][]{madhu2014,drummond2019}.

However, most of these efforts have, thus far, assumed a pre-defined ice abundance for the pebbles, which remains unchanged from the beginning of simulations until the pebbles are accreted onto the forming planet. An exception is the desorption of volatile ices, which is assumed to happen when the disk temperature around the pebbles reaches a certain level, corresponding to an icy species desorbing into the gas-phase.

The work by \citet{boothilee2019}, did utilise a chemical kinetics network through the \textsc{Krome}-package \citep{grassi2014} to model chemical evolution along with grain growth, although they only considered hydrogenation reactions for ices, and not two-body ice reactions via the Langmuir-Hinchelwood-mechanism \citep[see, e.g.][]{hasegawa1993,herbst2021review}. \citet{krijtbosman2020} used a dynamical model setup for grain evolution, including settling from upper disk layers to the midplane, grain growth and grain drift. They did account for ice chemistry, but to simplify the treatment of chemical evolution the chemical network was limited compared to other networks in use (they considered fewer chemical species). From the perspective of chemical modelling, studies like \citet{eistrup2016,eistrup2018,yu2017} utilised large chemical networks to model chemical evolving in disk midplanes, but usually assumed constant grains sizes throughout the chemical evolution (typically 0.1$\mu$m), and did not consider material mixing between different regions of the disk.

More recently, \citet{gavino2021} utilised the 3-phase \textsc{Nautilus}-code \citep[][]{ruaud2016} and a grain population with multiple grain sizes (constant in time) to model the evolution of volatile ices in a disk, accounting for both the radial and the vertical structure of the disk. Furthermore, \citet{clepper2022graingrowth} modelled chemical evolution during grain growth in the midplane, including the effect of turbulent mixing of material from the upper disk layers with the midplane. Both these studies with be further discussed in Section \ref{disc}.

This paper investigates the chemical evolution in volatile ices on growing grains, comparing grain populations with multiple different constant and evolving grain sizes, and utilising a full chemical kinetics code (2-phase). The goal is two-fold: 1. explore the effects of dust coagulation on ice abundances, and 2. compare how chemical evolution during realistic grain growth compares to chemical evolution assuming assuming simplified grain choices. This will offer a perspective on the level of sophistication needed or desired when implementing a treatment of grain growth into chemical kinetics codes, to account for grain growth during chemical evolution. This, in turn, will aid to a better understanding of the actual icy compositions of the resulting pebbles, and will provide a better basis of making the connection between exoplanet atmospheres to be characterised with \emph{JWST}, and protoplanetary disk midplane evolution.

\section{Methods}
\label{methods}
The work presented here implements a treatment of grain growth into a comprehensive chemical kinetics code. This approach is novel and necessary for understanding chemical evolution in the more realistic context of growing gains.

The code was modified for this work in order to account for evolving grain sizes, i.e. grain growth. While there are several effects leading to and resulting from grain growth (e.g., dust settling in the disk midplane, radial drift, and changes to the disk temperature structure and radiation field as opacities change) we focus on one effect of dust growth, namely the removal of dust surface area available for chemical ice reactions. The surface area on dust grains are important for the interactions between gas-phase species and grain-surfaces, because of the amount of area dust grains provide facilitating sticking of gas-phase species, as well as reaction between icy species.

Most chemical kinetics codes for modeling of interstellar and protoplanetary disk astrochemistry assume spherical grains with fixed sizes \citep[usually $R_{\rm grain}=0.1 \mu$m, see e.g.][]{fogel2011,walsh2015,eistrup2016}. However, there is both theoretical work and observational evidence \citep[see e.g.][]{testi2014,birnstiel2018_dsharp} to suggest that dust grains grow under conditions found in a disk. It is therefore crucial to develop a better understanding of the effects that grain growth may have on the efficiency of chemical evolution in disk midplanes.
\subsection{Set-up}

\subsection{Dust coagulation}\label{sec:dust_coagulation}
As a starting point for our simulations, we perform local dust coagulation calculations at the disk midplane at 4 radial locations: 1.5 AU, 5 AU, 20 AU and 30 AU. The temperatures of gas and grains (assumed thermally coupled), gas number densities and ionisation rates for each considered radius are indicated in Table \ref{grain_sizes}.

Assuming all grains start with a size of $0.1\mathrm{~\mu m}$, and maintaining a fixed total solids-to-gas ratio of 0.01, we use a representative particle method to calculate how successive grain-grain collisions alter the local dust size distribution. This method involves calculating collision rates between pairs of representative particles, making use of random numbers to determine which particles will be involved in the next collision event, and when that collision takes place \citep[see e.g.,][for more details]{zsom2008,drazkowska2014,krijt2016_dustgrowth}. When computing  collision rates we include Brownian motion, differential drift, and turbulence as sources for relative velocities between pairs of particles \citep[following e.g.,][]{birnstiel2016}, employing an $\alpha_T$ prescription for the turbulence and setting $\alpha_T=10^{-3}$ when calculating the contribution from turbulent motions using the formulation of \citep{ormelcuzzi2007}. We consider perfect sticking and catastrophic fragmentation to be the dominant collision outcomes, with fragmentation occurring above a fixed threshold velocity of $v_\mathrm{frag} = 10\mathrm{~m/s}$ as often taken to be appropriate for ice-covered grains \citep{birnstiel2016}. Grain charging is not considered when determining collision rates or outcomes.

The outcome of the simulations are evolving sizes for the representative particles that can be combined to produce size distributions or other derived quantities. We capture the evolving size distributions in Fig.~\ref{grain_evol} by showing the time evolution of two characterising sizes, the \emph{mass-dominating} grain size $R^m$ (dashed) and \emph{surface-area-dominating} grain size $R^A$ (solid), defined as
\begin{equation}
    R^m = \frac{\sum_i m_i R_i }{\sum_i m_i};~~~ R^A = \frac{\sum_i A_i R_i }{\sum_i A_i},
\end{equation}
where $R_i$, $m_i = (4/3)\pi R_i^3 \rho$ and $A_i= 4 \pi R_i^2$ represent the radius, mass and surface area of aggregate $i$ (assuming a spherical and compact structure), and $\rho=2.6\mathrm{~g/cm^3}$ is the material density.

At each location, we clearly see the effects of both coagulation and fragmentation. Initially, pairs of small, well-coupled grains collide at low velocities, allowing them to stick and grow, leading to an increasing $R^m$ and $R^A$. The difference between $R^m$ and $R^A$ is small during this phase, indicating that the underlying dust size distribution is quite narrow. As the largest grains begin to approach sizes of ${\approx}10^{-1}\mathrm{~cm}$, however, collision velocities become comparable to the fragmentation velocity, and the ensuing disruptive events lead to the creation of small fragments, the introduction of which quickly decreases $R^A$. Further growth of the largest aggregates stagnates as the fragmentation limit is reached, and $R^m$ begins to level of. After some time, a steady state is reached in which the removal and creation of grains are balances at every size, and both $R^m$ and $R^A$ are constant in time. Typically in protoplanetary disk environments these fragmentation-limited steady-state distributions are such that most of the mass is contained in large aggregates while smaller ones dominate the number and surface area \citep[see][]{Birnstiel2011}. Indeed, the final difference between $R^m$ and $R^A$ is around 2 orders of magnitude, highlighting the broad nature of the final equilibrium situation. Comparing different radial locations in Fig.~\ref{grain_evol}, we see that towards \emph{larger} disk radii: coagulation timescales become longer, while maximum and equilibrium grain sizes decrease. The behavior described above is in agreement with the established picture of (local) dust coagulation \citep[as reviewed for example in][]{birnstiel2016}. We note that our local coagulation calculations do not include radial drift or other forms of material transport between different disk regions; we discuss these limitations further in Sect.~\ref{sec:dust_dynamics}.

In the rest of this paper, two grain size cases from Fig. \ref{grain_evol} are used, for each radius. The first is the case in which the representative grain size $R^A$ evolves for each step in the chemical evolution. Since, in this case, $R^A$ varies for each time step, this grain size case will be denote ``vary''. The other case is that in which the grain coagulation and fragmentation dynamic has led to a steady-state, or ``converged'' value for $R^A$, which is the flattening out of the solid profiles for $R^A$ in Fig. \ref{grain_evol}, which happens for all four radii, after the value for $R^A$ has peaked. Chemical evolution models assuming this steady-state, or converged, grain size $R^A$, which is derived from grain evolution models, will henceforth be referred to as ``conv''-models.



\subsection{Volatile chemistry}
The chemical kinetics code previously used in works of \citet{walsh2015,eistrup2016,eistrup2018} was utilised. This code includes gas-phase reactions, gas-grain interactions and grain-surface reactions, including two-body ice reactions and hydrogenation reactions. The chemical kinetics code from \citet{walsh2015} was modified to enable variations in the grain sizes. This was done by changing the radius $R_{\rm grain}$ of dust grains from a global constant in the code to being a time variable where a new value $R_{\rm grain}$ is read for every time step.

\begin{figure}
    \centering
    \includegraphics[width=.45\textwidth]{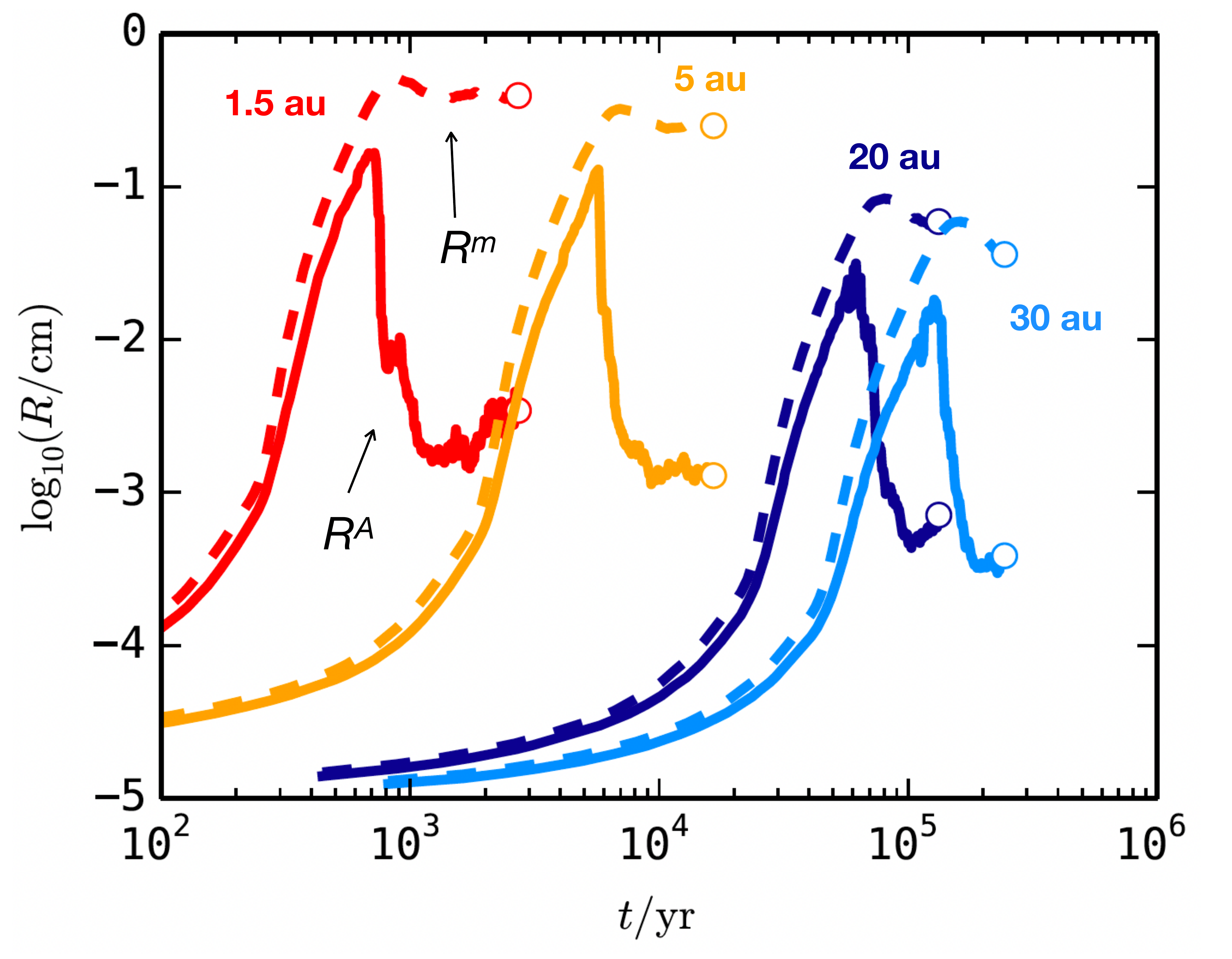}
    \caption{Evolution of characteristic grain size over time at four heliocentric distances. Curves display the mass-dominating size $R^m$ (dashed) and surface-area-dominating grain size $R^A$ (solid). After an initial phase of efficient coagulation (marked by $R^m$ and $R^A$ increasing similarly), the creation of fragments in disruptive collisions acts to lower $R^A$ and halt growth altogether as a steady state is reached towards the end of the local simulations (shown by open white circles).}
    \label{grain_evol}
\end{figure}

As seen in Fig. \ref{grain_evol} the characteristics (maximum grain size, $R^m$ and representative grain size $R^A$) change over time. To simplify the implementation of the dust evolution into the chemical kinetics code, an effective grain size representative of the average grain-surface area per grain $R^A$ across the whole dust size distribution is derived for each time step. Using $R^A$, and changing its value as a function of time allows the actual grain surface area of the entire dust grain population to be represented by a single grain size for each time step.

The implementation into the code was done as follows: at the first time step, the initial grain size value $R_{\rm grain,t=0}$ is used to calculate chemical reaction rates of effects involving collisions between gas species (atoms, molecules and electrons) and grains, effects for which the size of the grains matter. In addition to using the initial grain size, the number density of grains per H$_2$ $N(\rm grains)$ is also used. In previous astrochemical models of protoplanetary disk midplanes using fixed-size spherical grains of 0.1$\mu$m \citep[see e.g.][]{eistrup2016,eistrup2018}, the grain number density was set to $N(\rm grains)=1.3\times 10^{-12}$ per H$_2$. That grain size and grain number density correspond to a gas-to-dust mass ratio of 100. In this work, the grain growth is assumed to happen locally (fixed disk midplane radius), thus without gain or loss of material. In order to conserve the total mass of the dust grains, the grain number density $N(\rm grains)$ will evolve from one time step to the next, corresponding to the change in $R_{\rm grain}$, such that the total dust mass is kept constant. This is achieved by scaling $N(\rm grains)$ for all time steps except the first, in the following way:
\begin{equation}
\begin{split}
    N_{t=n}(\mathrm{grains})&=N_{t=n-1}(\mathrm{grains})\frac{m_{t=n-1}(\rm grains)}{m_{t=n}(\rm grains)}\\
    &=N_{t=n-1}(\mathrm{grains})\left(\frac{R_{t=n-1}(\rm grains)}{R_{t=n}(\rm grains)}\right)^3,
    \label{grain_rd}
\end{split}
\end{equation}
where $m$(grains)=$\frac{4}{3}\pi R^3\rho$(grains) is the mass of a single grain of radius $R$ (where $\rho$(grains) is the mass density of the grains [g cm$^{-3}$]), and $t=n$ is the $n$th timestep in the grain size evolution. If the initial grain size differs from 0.1$\mu$m, then the initial grain number density $N_{t=0}$(grains) will be adjusted from the value 1.3$\times 10^{-12}$ accordingly, so as to ensure the dust mass conservation.

Given the nature of grain \emph{growth}, namely that the dust grains collide, coagulate and grow larger in size, it can be inferred from Eq. \ref{grain_rd} that the grain number density will decrease as a function of time. However, for the grain evolution that is considered here, which includes fragmentation of larger dust grain to form smaller grains, smaller grains are replenished and thereby a population of small grains is sustained a later stages of grain evolution. Although smaller grains account for a small amount of the total dust mass, they contribute an appreciable amount of surface area to the whole dust population. For this reason, as also seen in Fig. \ref{grain_evol}, the effective grain size $R^A$ actually peaks and subsequently decreases before converging. This evolution of the effective grain size is accounted for by means of the implementation into the chemical kinetics code as outlined in Eq. \ref{grain_rd}.

The principle behind the treatment of grain sizes in this study is the following: in the actual grain size distribution, it is assumed that all grains are spherical, that they have different sizes, and for each time step during evolution, there is a certain number of them. In the treatment here, the formal simplification is made that all grains are one size for each time step (solid profiles in Fig. \ref{grain_evol}). This one size per time step is calculated such that this homogeneously-sized population of grains features the same total grain surface area as the original distribution with different sizes, assuming that both the cases with homogeneously-sized grains and the actual grain distribution with different sizes have identical number densities for the same time step. This principle allows for an implementation of grain growth into the chemical kinetics code where the effective grain size is updated for each time step, as a representation of the surface-area-wise behavior of the actual grain distribution over time.

\begin{table*}
    \centering
    \begin{tabular}{l c c c l c c}
        Radius [AU] &Initial grain &Maximum grain &Converged (final) [cm]&$T$&$N$(\ce{H2})&$\xi$\\
        & size [cm]&size $R^A$ [cm]&grain size $R^A$&[K]&[cm$^{-3}$]&[s$^{-1}$]\\
        &&(time reached [yr])&(time reached [yr])&&\\
        \hline\hline\\
        1.5 &$10^{-5}$&1.61$\times10^{-1}$ (7.18$\times10^{2}$) &3.43$\times10^{-3}$ (2.72$\times10^{3}$)&120 &7.2$\times10^{12}$&7.2$\times10^{-18}$\\
        5 &$10^{-5}$ &1.29$\times10^{-1}$ (5.70$\times10^{3}$)&1.29$\times10^{-3}$ (1.65$\times10^{4}$)&57&9.2$\times10^{10}$&9.7$\times10^{-18}$\\
        20 &$10^{-5}$ &3.14$\times10^{-2}$ (6.17$\times10^{4}$)&7.11$\times10^{-4}$ (1.33$\times10^{5}$)&25 &1.5$\times10^{10}$ &1.0$\times10^{-17}$\\
        30 &$10^{-5}$ &1.82$\times10^{-2}$ (1.27$\times10^{5}$)&3.85$\times10^{-4}$ (2.44$\times10^{5}$)&19.5 &5.4$\times10^{9}$&1.0$\times10^{-17}$\\ \\\hline
    \end{tabular}
    \caption{Maximum and converged grains sizes $R^A$ for four radii depicted in Fig. \ref{grain_sizes} (solid profiles), and times during evolution where these are reached. Midplane temperature $T$, hydrogen volume density $N$(\ce{H2}) and ionisation rate $\xi$ are also noted for reference.}
    \label{grain_sizes}
\end{table*}

It is noted that, when modelling chemical evolution using various grain sizes at a local radius in a disk midplane without gain or loss of material, it is important to conserve the grain mass density $N_{\mathrm{grains}}$ per hydrogen atom. That is, as grains grow larger, the number of grains will decrease.

When assuming 0.1$\mu$m grains, a gas/dust ratio of 100 and a silicate grain mass volume density (3g/cm$^{3}$), the grain number density per hydrogen atom is $N_{\mathrm{grains}}$(0.1$\mu$m)=1.3$\times10^{-12}$. When assuming larger (constant-sized) grains of size $R_{\mathrm{choice}}$, the grain number density will be constant, but smaller than 1.3$\times10^{-12}$, assuming the other factors stay constant. This will be accounted for, before running the chemistry.

This scaling means that for each factor of 10 the grains are assumed to increase in size (1$\mu$m versus 10$\mu$m), the number density of grains per hydrogen atom decreases with a factor of 1000. For the grain size case in which the grains are growing for each timestep in the chemical evolution, but where the grains start out with sizes of $R_{\mathrm{grains}}$=0.1$\mu$m, the adjustment for the grain number density happens automatically in the chemical kinetics code, as a consequence of reading in a new grain size.

Dust grains are also important for the balance and exchange of electric charge in disk midplanes. Dust grains can become electrically charged by accreting electrons, and negatively charged grains can, in turn, neutralised gas-phase cations through collisions. Previous chemical models \citep{eistrup2016,eistrup2018} assumed grain number conservation, and were initialised with only neutral grains. However, as these models evolved the chemistry, a fraction of the total grains could become singly negatively charged, and the sum of the number of negatively charged grains and the neutral grains would equate to the initial number of neutral grains.

Like those previous works, this work only assumes neutral and singly negatively charged grains (Gr$^0$ and Gr$^-$, respectively). The chemical models are initialised in a charge-neutral state (in this case, $N_{t=0}$(Gr$^-$)=0, $N_{t=0}$(e$^-$)=0, $N_{t=0}$(cations)=0 and $N_{t=0}$(anions)=0). For a given evolution time step after the $t_0$, featuring a non-zero electron abundance, a fraction of the grains will capture electrons and thereby become negatively charged. By using a chemical kinetics code with constant dust mass and dust grain sizes, the sum of the number densities of negatively-charged ($N_{t}$(Gr$^-$)) and neutral grains ($N_{t}$(Gr$^0$)) at time $t$ will simply equate to the initial neutral grain number density.

However, when including grain growth with dust grain mass conservation, as grains are growing the number of grains will decrease (see Eq. \ref{grain_rd}). If, at the beginning of each time step, the number densities of neutral and charged grains are scaled according to Eq. \ref{grain_rd}, this results is a net change of charge held in grains (net loss, if grains are growing). In order to conserve charge, this change in charge in grains need to be accounted for elsewhere. There are several possible solutions to this issue: \emph{1} allowing individual grains to carry more than one unit of net charge when they grow, \emph{2} adjusting the efficiency of grain neutralisation reactions from collisions with gas-phase cations, which in turn may affect the cation abundances and charge balance in the gas, and \emph{3} adding/subtracting any gain or loss of grain charge to the electron abundance. The third possibility is the most straight-forward to implement, and it may affect the gas-phase cation abundances in a way similar to the second option, as an increase/decrease in electron abundances will affect the rates of cation-electron recombination reactions.

It noted that, in the chemical kinetics code, the rates for both electron capture by grains, and grain-cation recombination are functions of grain radius, the effects on these rates are therefore accounted for when changing grain sizes. It is further noted that for calculating the grain-cation recombination rates, the chemical kinetic code includes the rate enhancement prescribed by \citet{drainesutin1987}.

The second option requires a larger expansion of the chemical kinetics network, in order to enable negatively-charged grains to add their charges when they stick and grow together, as well as possible electron capture by grains that are already charged. Furthermore, as a destruction pathway for multiply-charged grains, full or partial neutralisation of these grains through charge-exchange reactions with cation in the gas-phase would need to be added to the chemical network, with one reaction for each cation. Such a more realistic expansion of the treatment of charged grains in the chemical network will be pursued in future work. In this work, the third option will be utilised to account for charge conservation.

Lastly, two different sets of initial abundances were utilised in this study: one featuring atomic abundances, and one featuring molecular abundances. The initial abundance sets for these two setups are identical to the abundances used in \citet{eistrup2016}, and we refer to Table 1 in that paper for more insights.

\section{Results}

In this section the results of the chemical modeling will be presented. Emphasis will be placed on highlighting the differences between chemical abundances for the nominal case with 0.1 $\mu$m-sized grains, and abundances for other grain size choices.

\begin{figure*}[h!]
    \centering
    \includegraphics[width=0.6\textwidth]{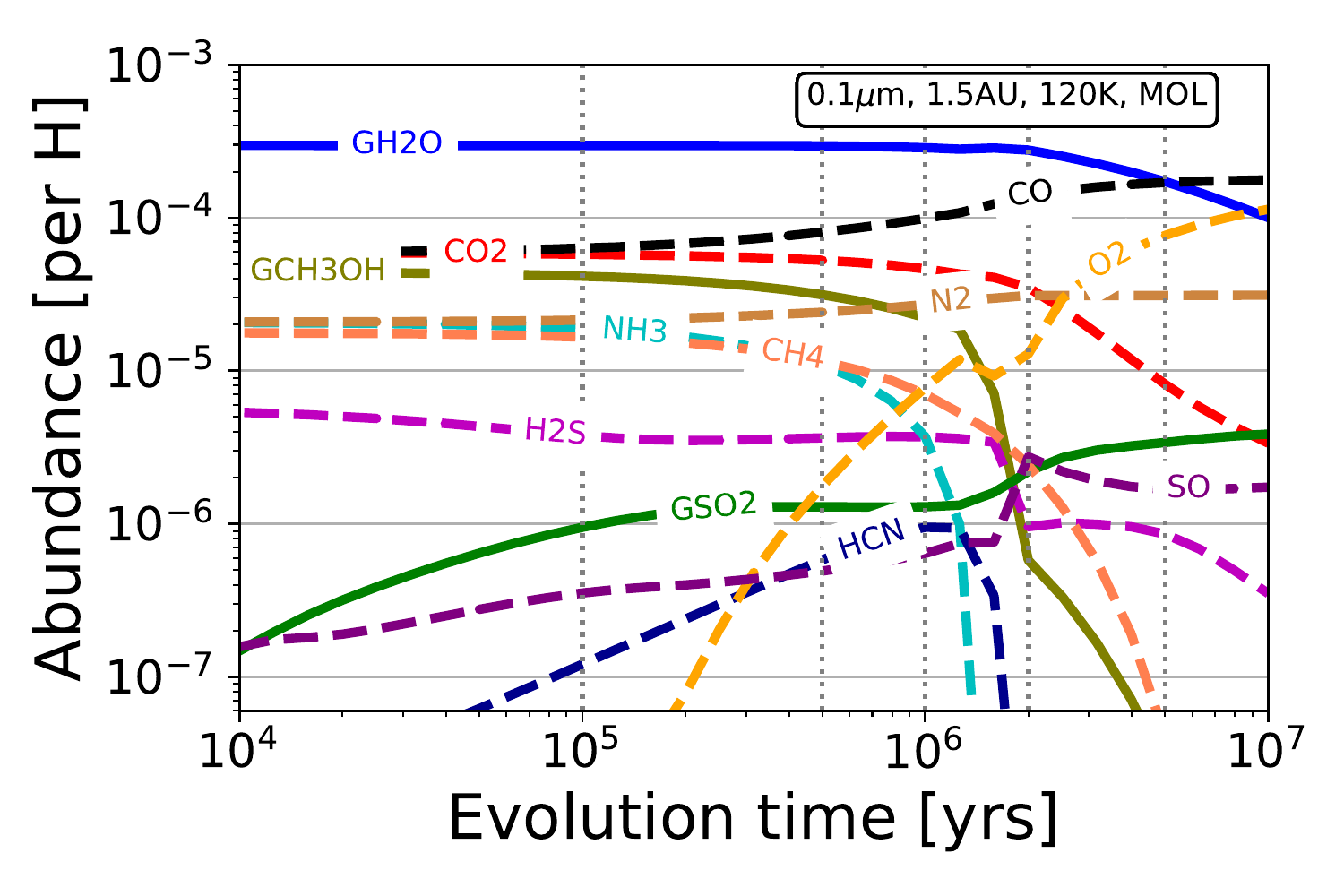}\\
    \includegraphics[width=1\textwidth]{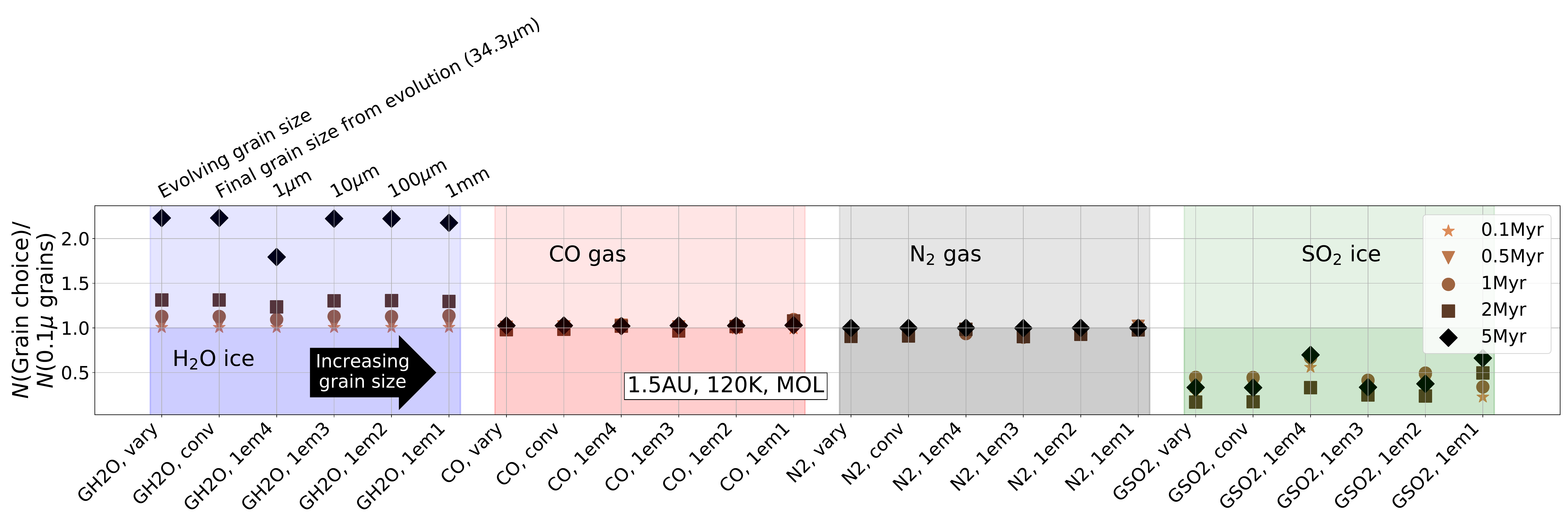}\\
    \includegraphics[width=1\textwidth]{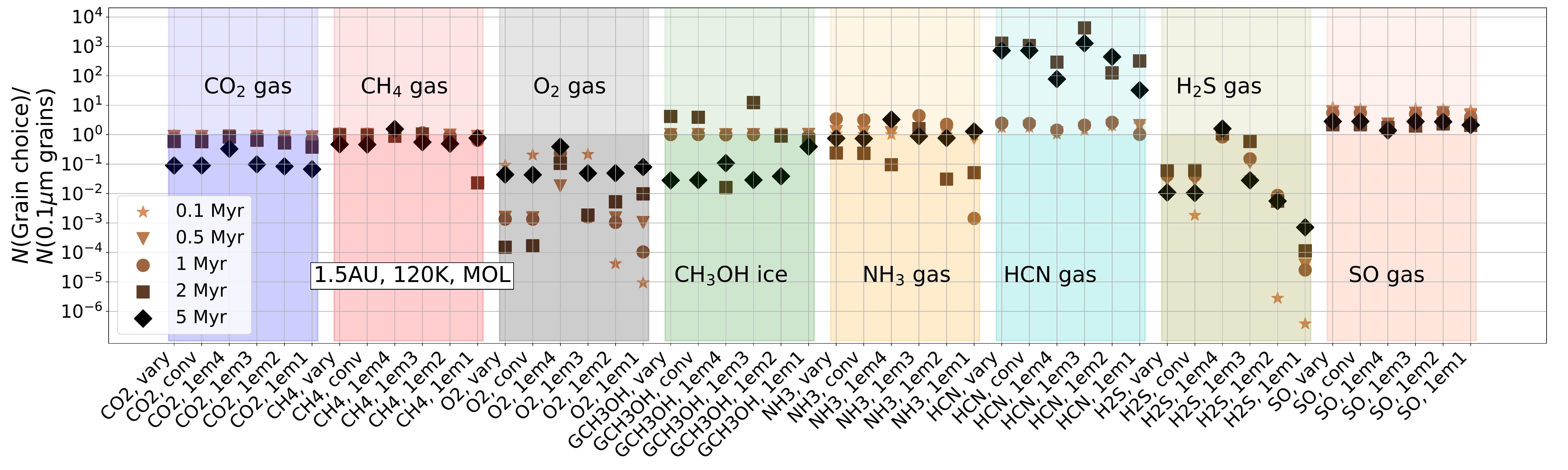}\\
 \caption{Top panel: evolution of chemical abundances at 1.5AU starting with molecular (``MOL'') abundances, for a constant grain size of $R_{\mathrm{grain}}=0.1\mu$m. Solid profiles are for ices (species names starting with ``G''). Dashed profile are for gases. Middle and bottom panels: time -and grain-choice-dependent abundances, relative to abundances assuming $R_{\mathrm{grain}}=0.1\mu$m.. The $y$-axes are linear for the middle panels, and logarithmic for the bottom ones. Vertical dotted lines in top panel indicate evolution times (0.1Myr, 0.5Myr, 1Myr, 2Myr and 5Myr) associated with markers in middle and bottom panels. The blue shaded area for \ce{H2O} ice in middle panel has extra annotation, to guide the reader: first vertical category is assuming an evolving grain size, growing with time (as annotated above the plot). The second category assumes the final (constant) grain size from the grain growth models, which here, at 1.5AU, is 34.3$\mu$m (see Table \ref{grain_sizes}). The following four categories represent log-spaced grain size increases (as also indicated with the black arrow, indicating increasing grain sizes going from categories three through six). This sequence of $x$-axis-categories is identical across all middle and bottom panels in Figs. \ref{1_5_mol}-\ref{30_atom} in this paper.}
 \label{1_5_mol}
\end{figure*}

\begin{figure*}[h!]
    \centering
    \includegraphics[width=0.6\textwidth]{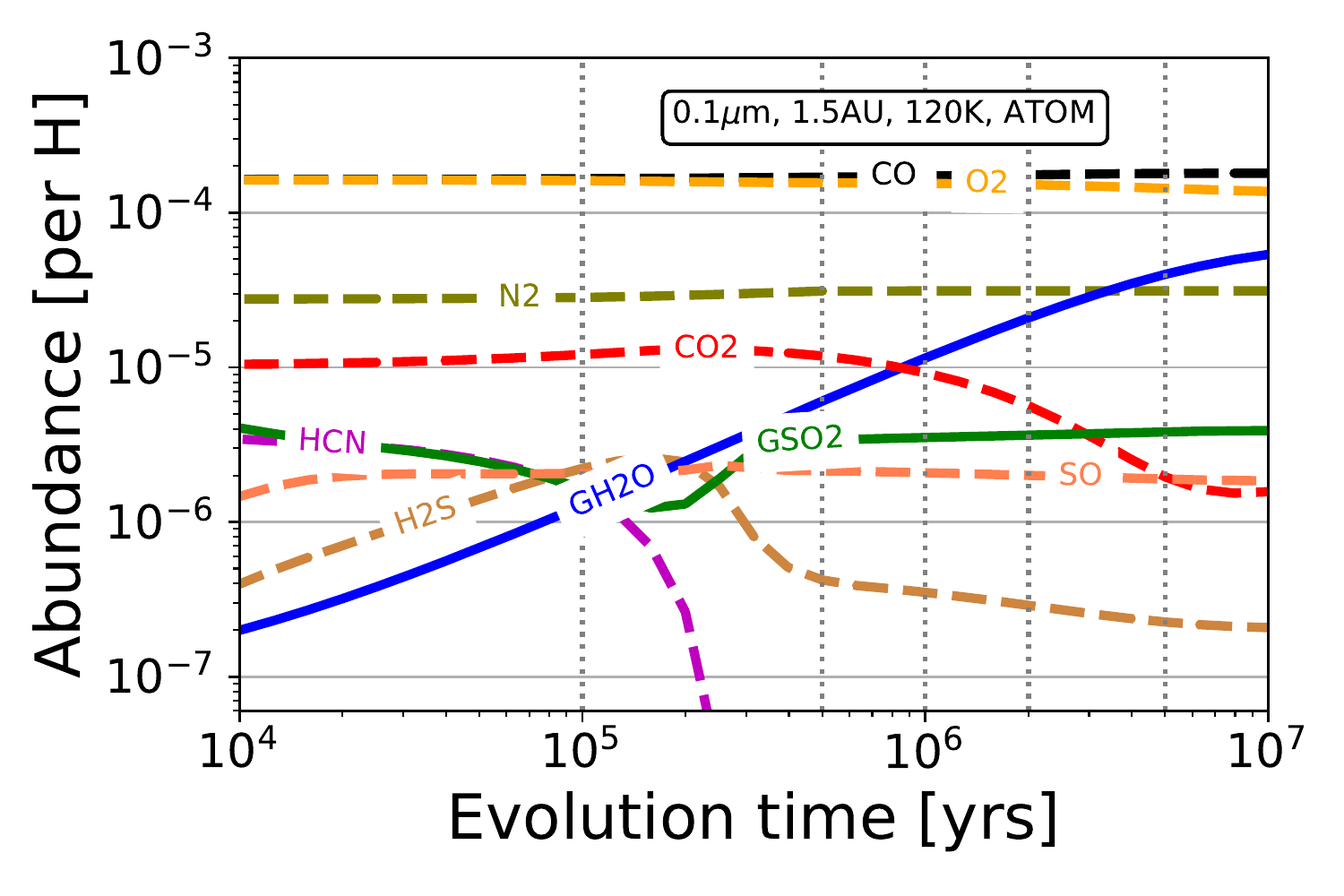}\\
    \includegraphics[width=1\textwidth]{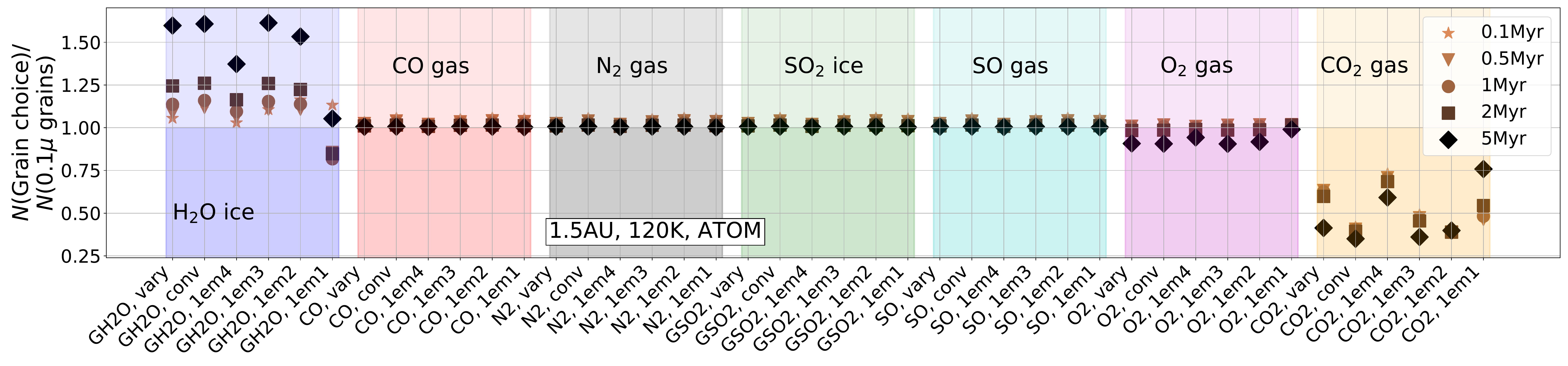}\\
    \includegraphics[width=1\textwidth]{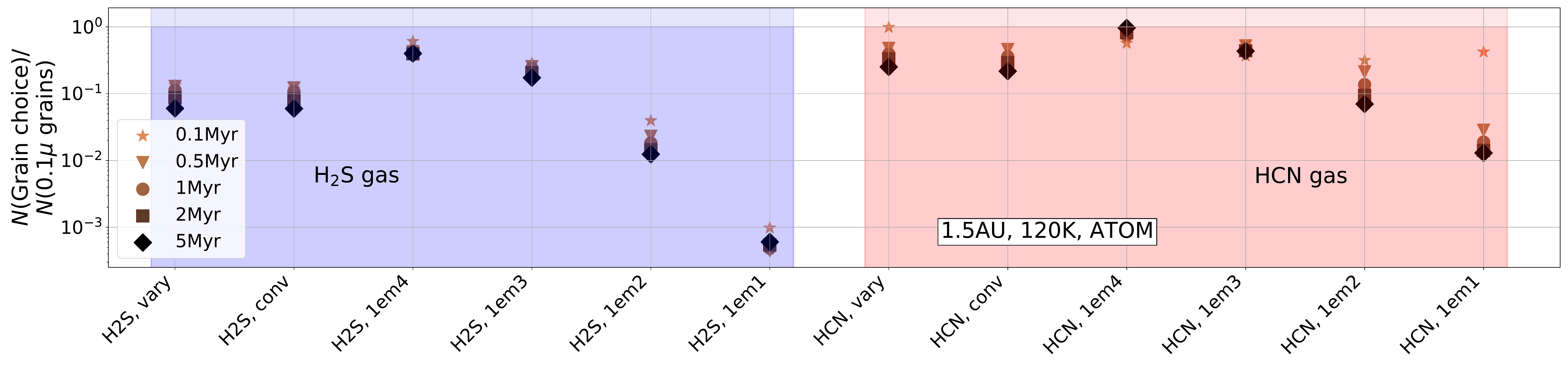}
    \caption{Similar to Fig. \ref{1_5_mol}, but for 1.5AU, and starting with atomic initial abundances. The converged (``conv'') grain size at 5AU is 34.3$\mu$m (see Table \ref{grain_sizes}). Please refer to caption of Fig. \ref{1_5_mol} for detailed description.}
    \label{1_5_atom}
\end{figure*}

\begin{figure*}
    \centering
    \includegraphics[width=.6\textwidth]{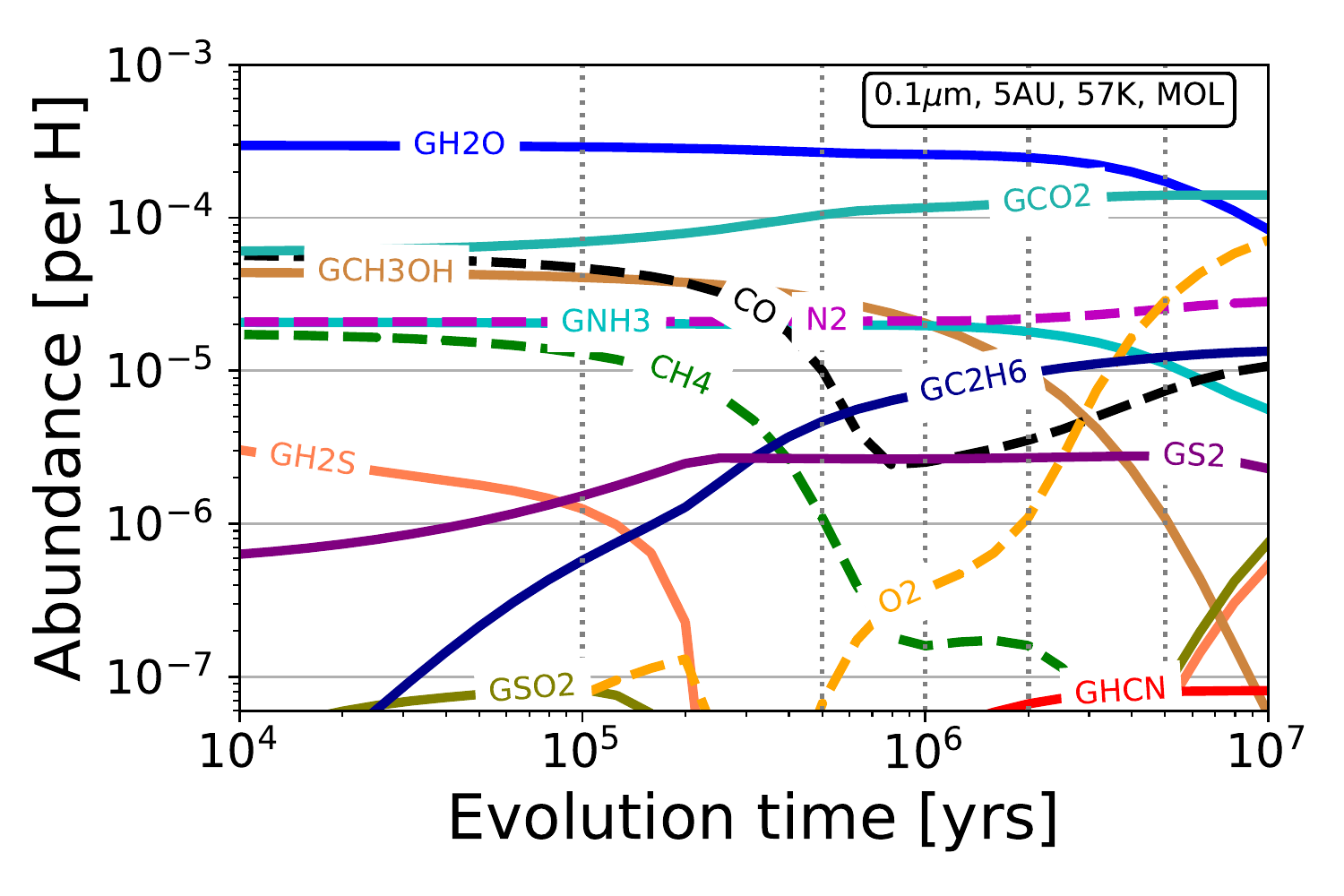}\\
    \includegraphics[width=1\textwidth]{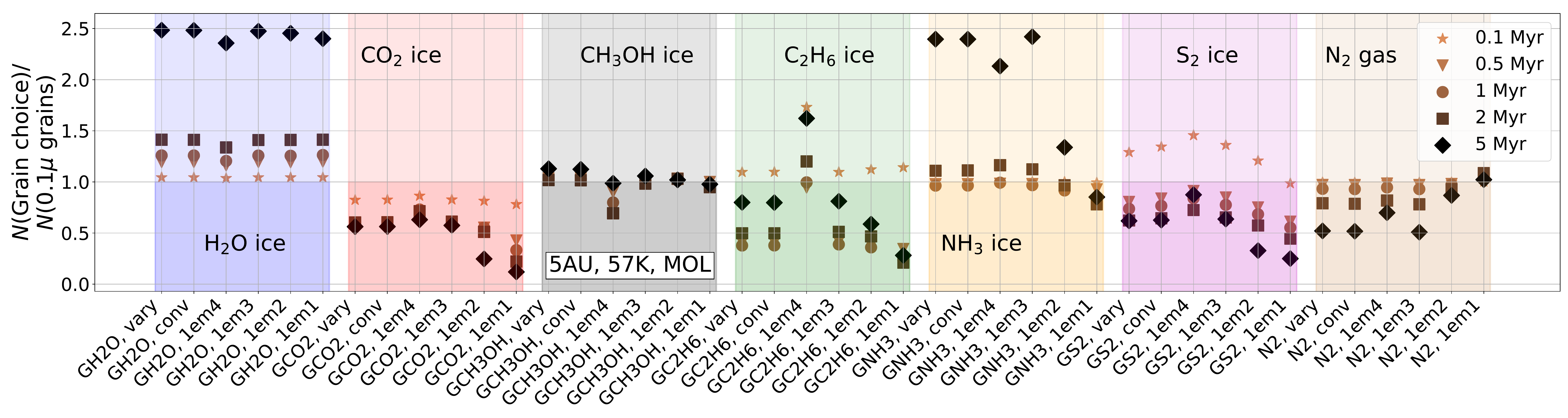}\\
    \includegraphics[width=1\textwidth]{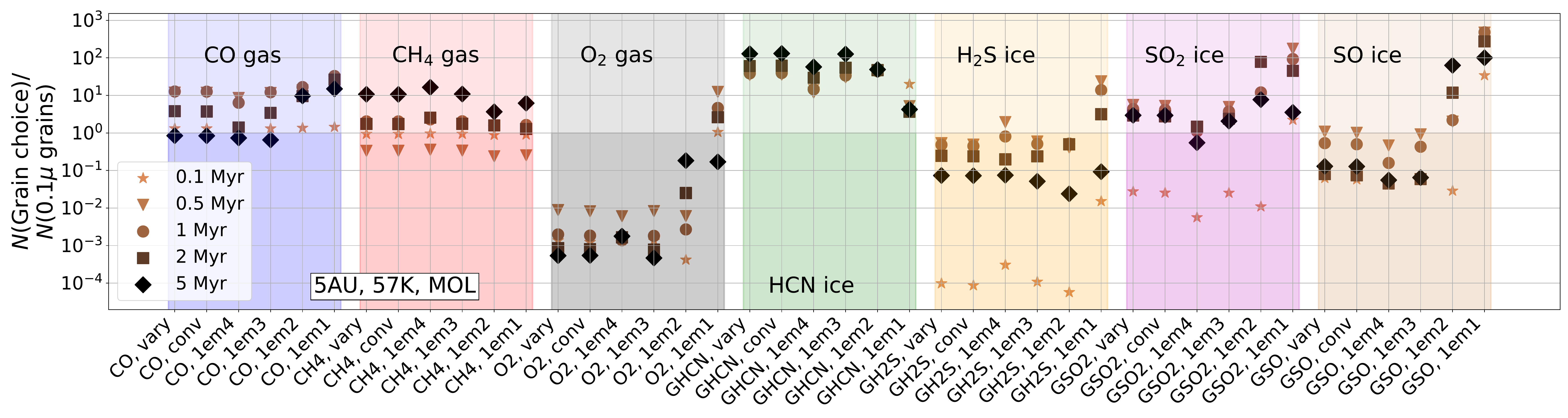}\\
    \caption{Similar to Fig. \ref{1_5_mol}, but for 5AU, and starting with molecular initial abundances. The converged (``conv'') grain size at 5AU is 12.9$\mu$m (see Table \ref{grain_sizes}). Please refer to caption of Fig. \ref{1_5_mol} for detailed description.}
    \label{5_mol}
\end{figure*}

\begin{figure*}
    \centering
    \includegraphics[width=.6\textwidth]{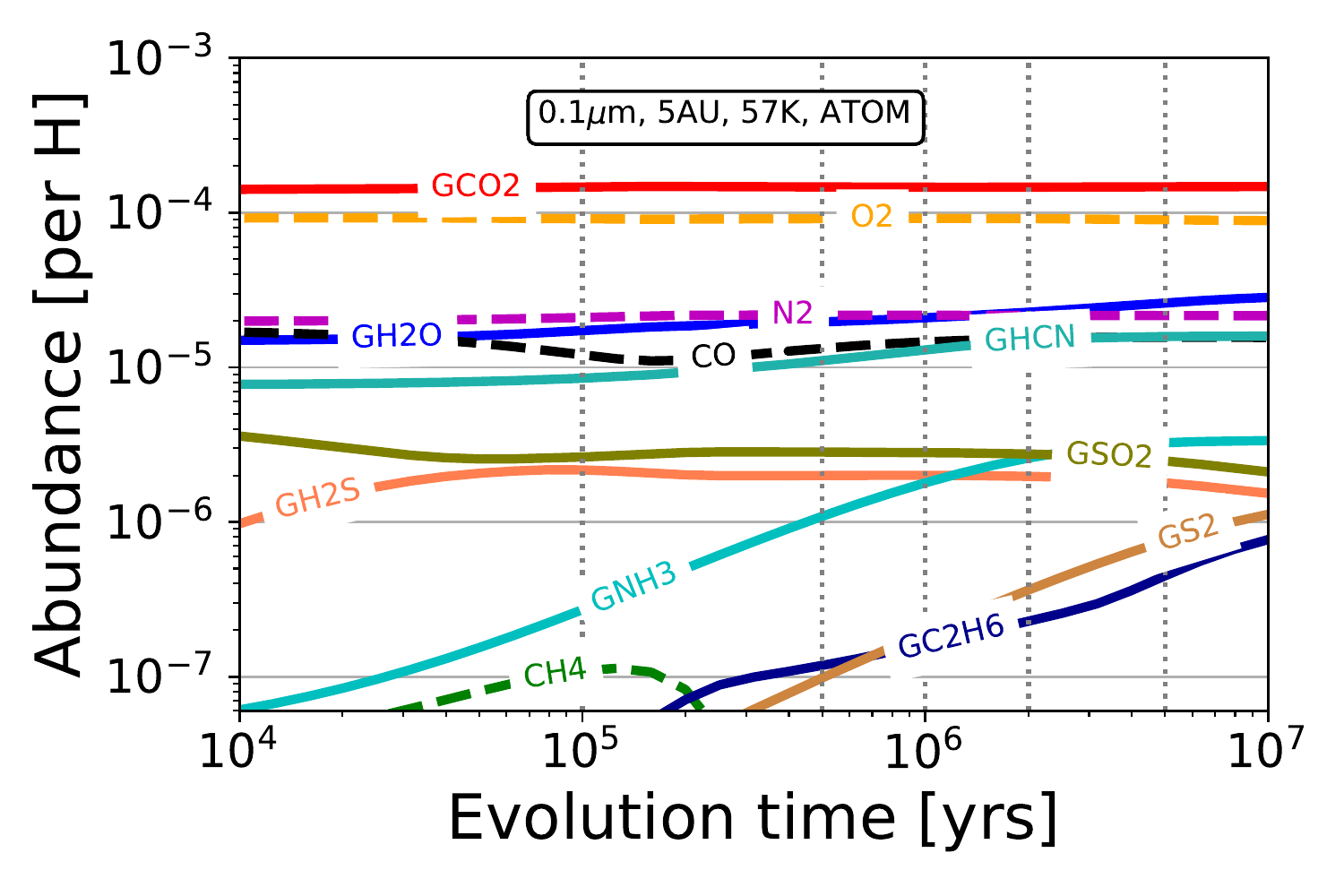}\\
    \includegraphics[width=1\textwidth]{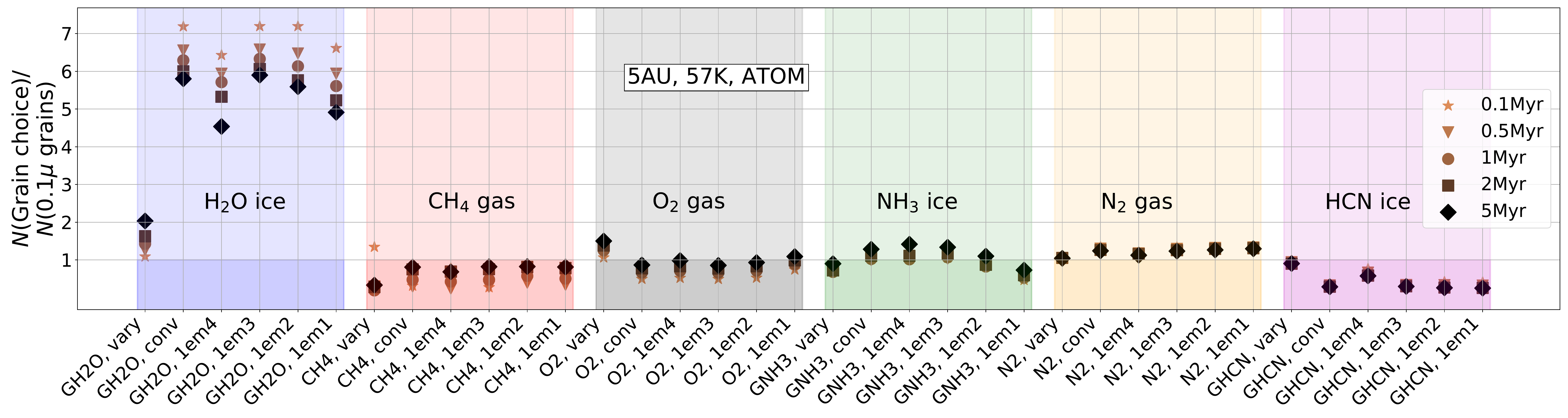}\\
    \includegraphics[width=1\textwidth]{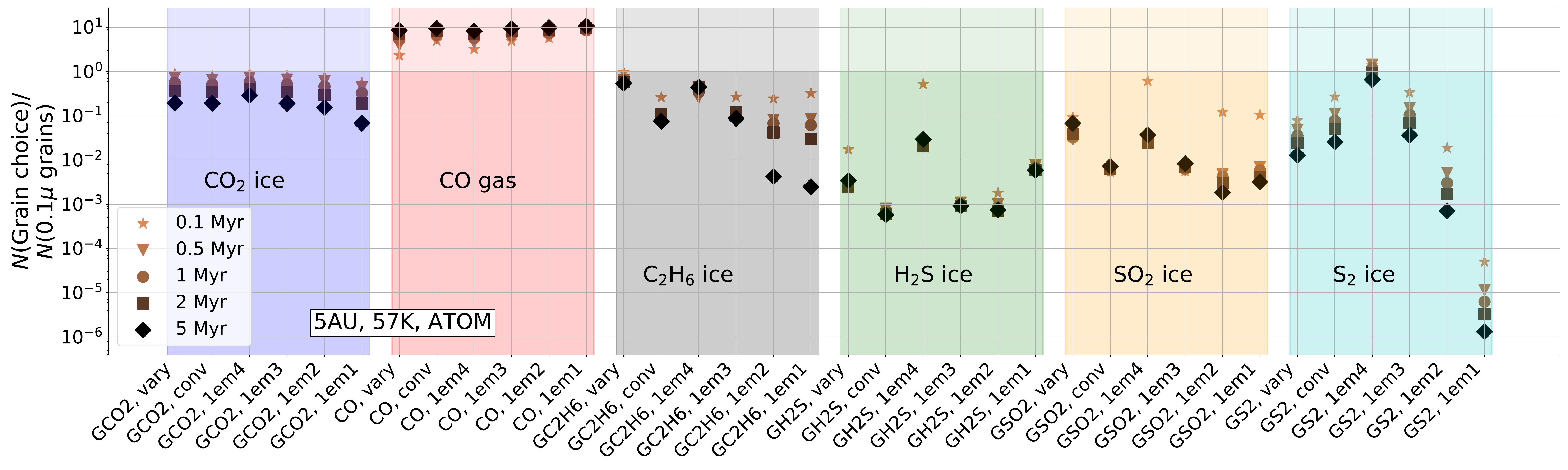}\\
    \caption{Similar to Fig. \ref{1_5_mol}, but for 5AU, and starting with atomic initial abundances. The converged (``conv'') grain size at 5AU is 12.9$\mu$m (see Table \ref{grain_sizes}). Please refer to caption of Fig. \ref{1_5_mol} for detailed description.}
    \label{5_atom}
\end{figure*}

\begin{figure*}
    \centering
    \includegraphics[width=.6\textwidth]{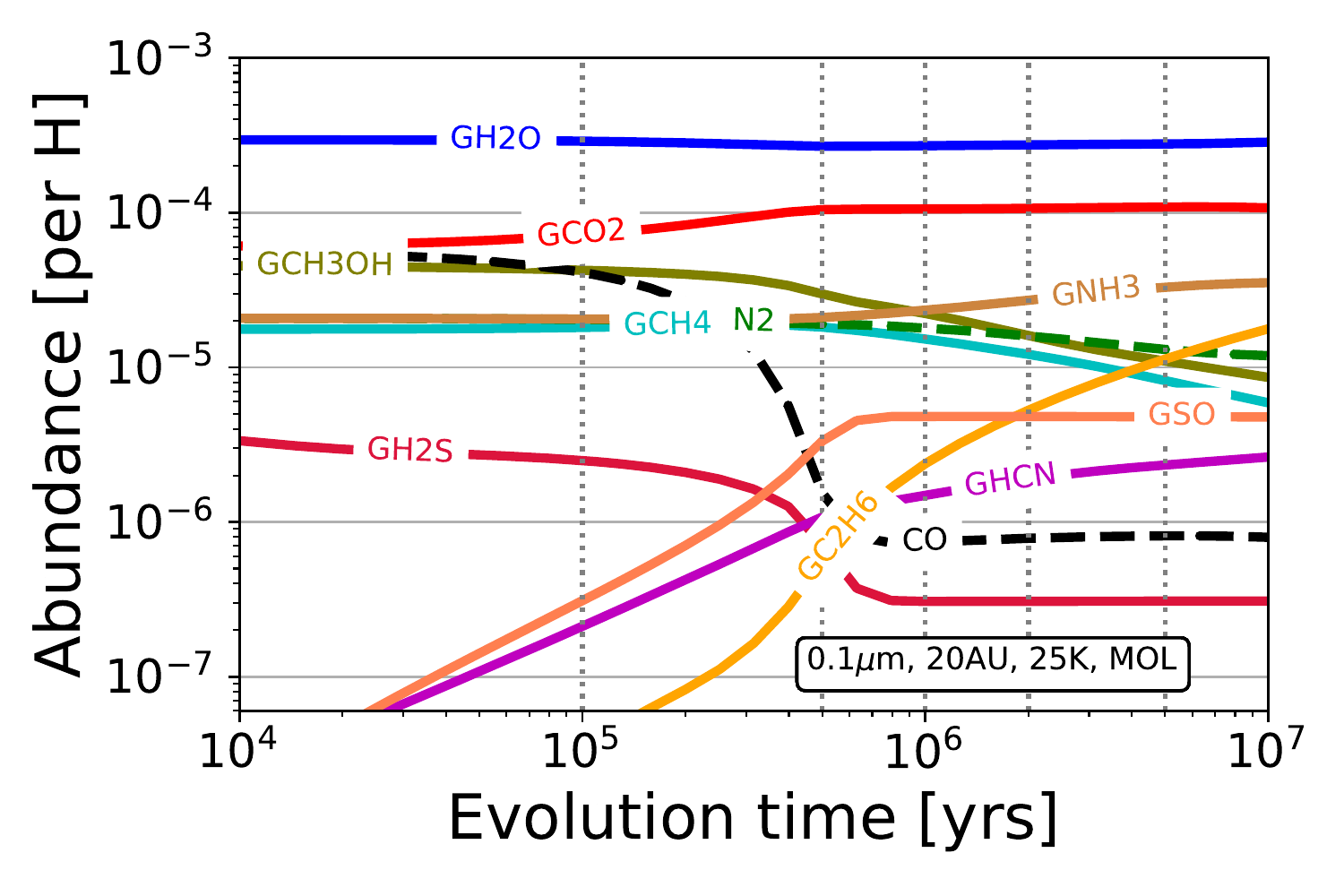}\\
    \includegraphics[width=1\textwidth]{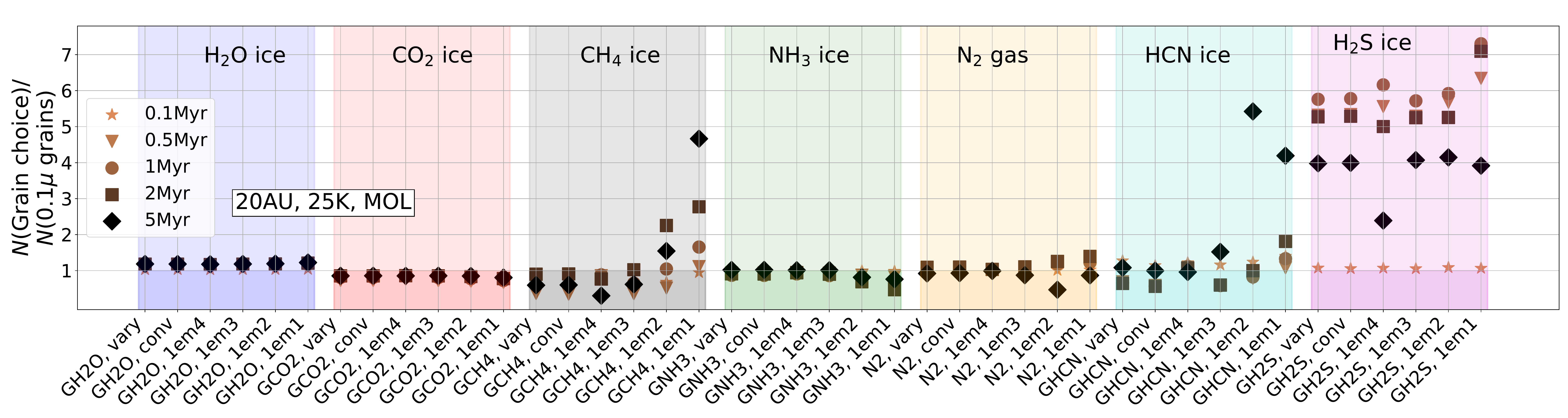}\\
    \includegraphics[width=1\textwidth]{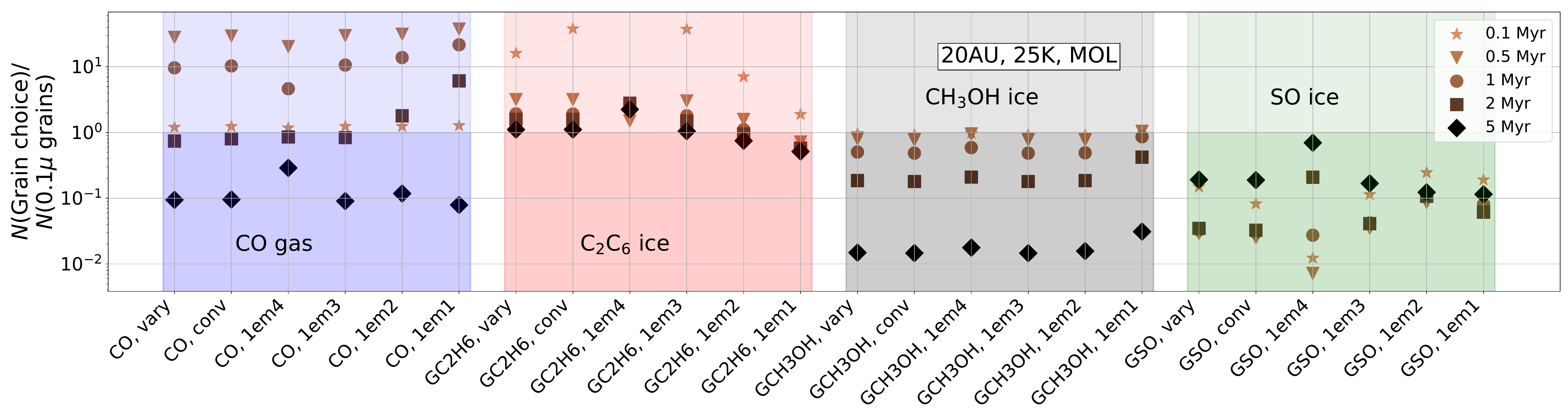}\\
    \caption{Similar to Fig. \ref{1_5_mol}, but for 20AU, and starting with molecular initial abundances. The converged (``conv'') grain size at 20AU is 7.11$\mu$m (see Table \ref{grain_sizes}). Please refer to caption of Fig. \ref{1_5_mol} for detailed description.}
    \label{20_mol}
\end{figure*}

\begin{figure*}
    \centering
    \includegraphics[width=.6\textwidth]{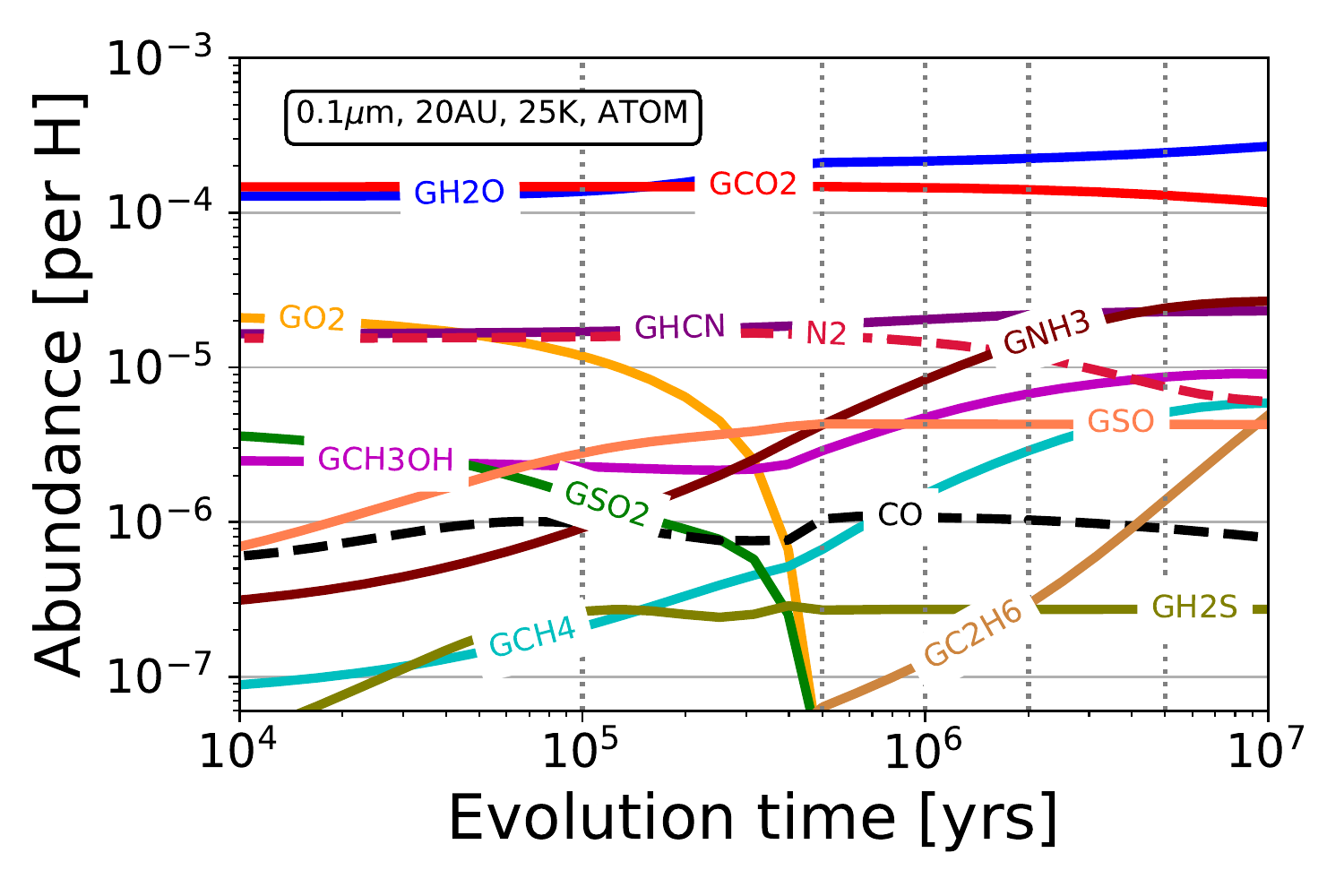}\\
    \includegraphics[width=1\textwidth]{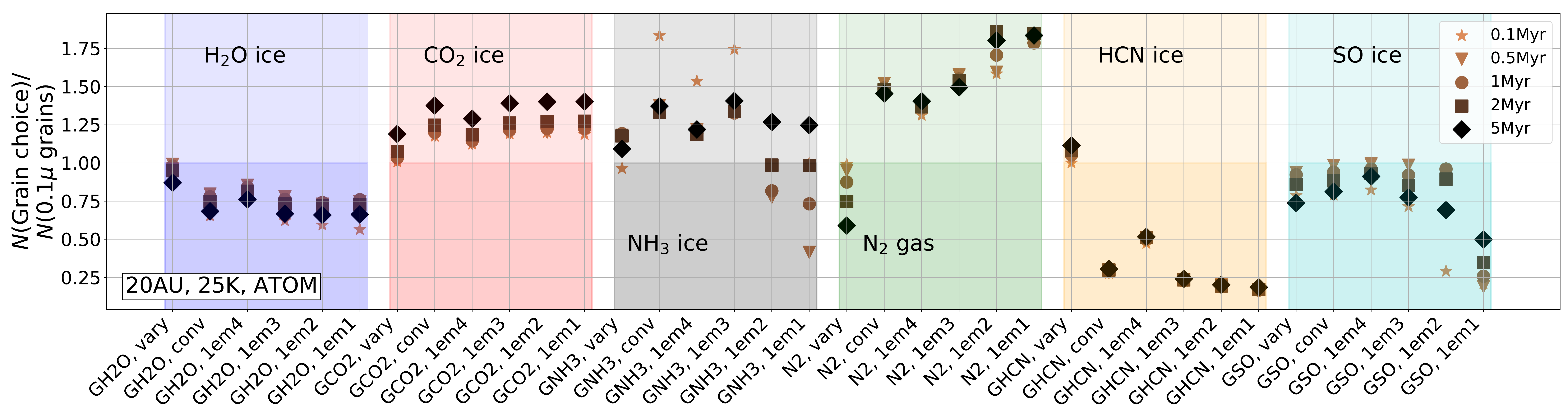}\\
    \includegraphics[width=1\textwidth]{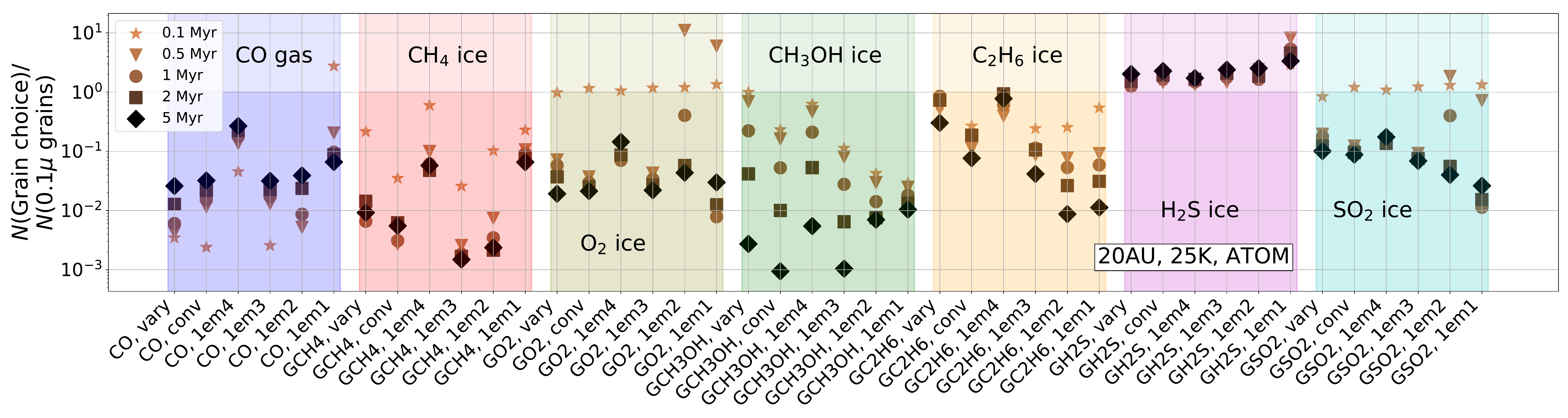}\\
    \caption{Similar to Fig. \ref{1_5_mol}, but for 20AU, and starting with atomic initial abundances. The converged (``conv'') grain size at 20AU is 7.11$\mu$m (see Table \ref{grain_sizes}). Please refer to caption of Fig. \ref{1_5_mol} for detailed description.}
    \label{20_atom}
\end{figure*}

\begin{figure*}
    \centering
    \includegraphics[width=.6\textwidth]{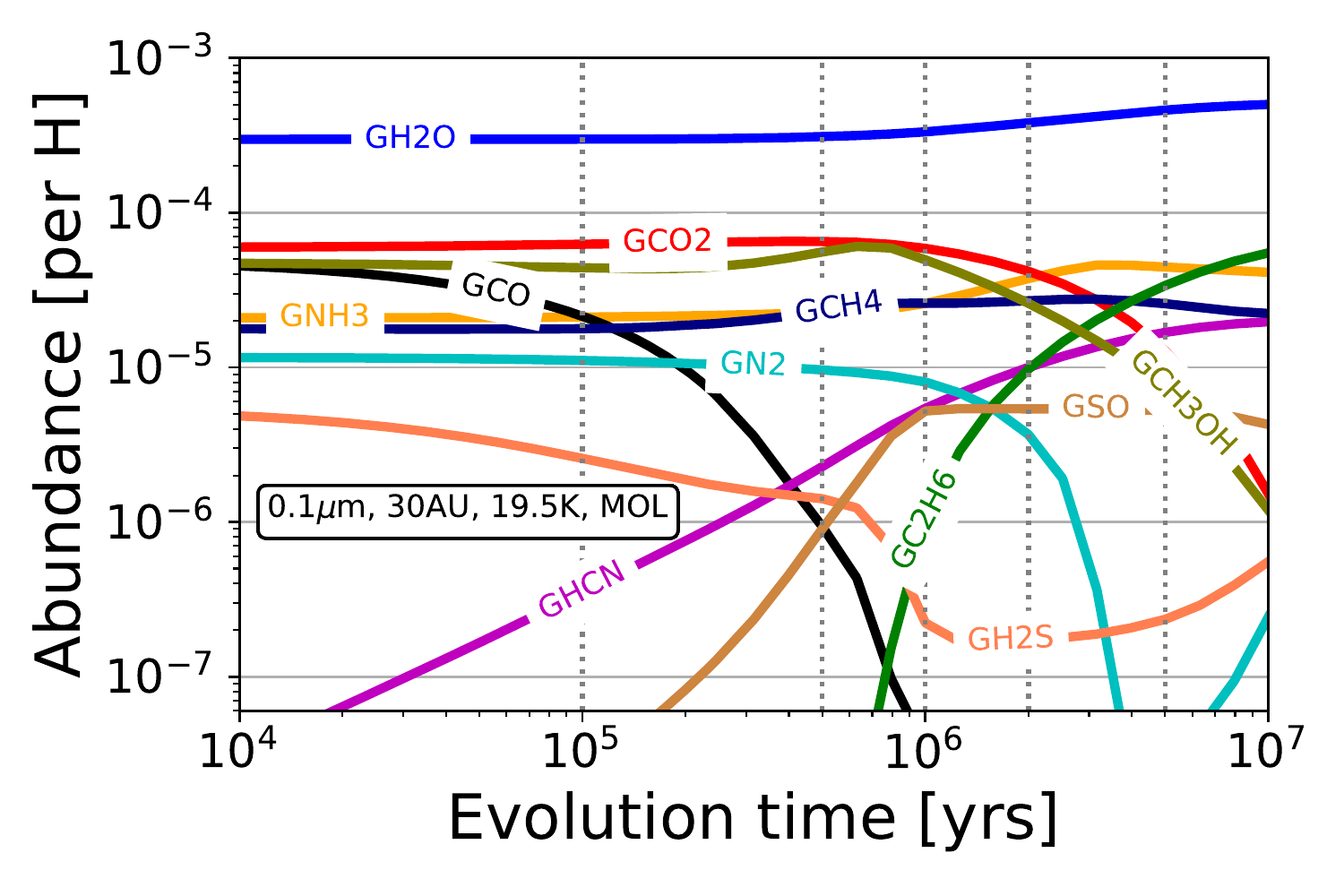}\\
    \includegraphics[width=1\textwidth]{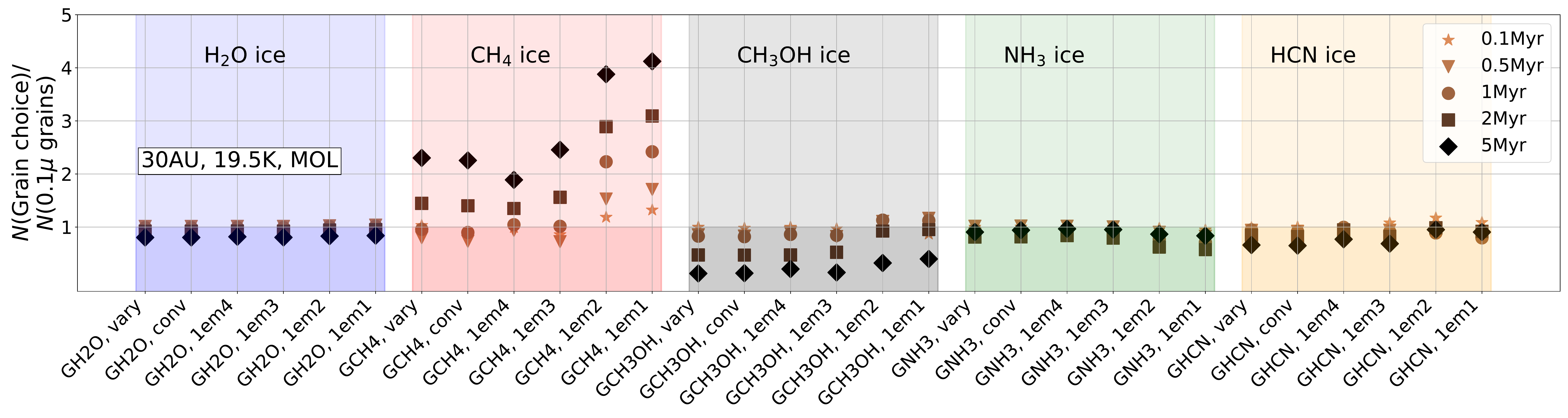}\\
    \includegraphics[width=1\textwidth]{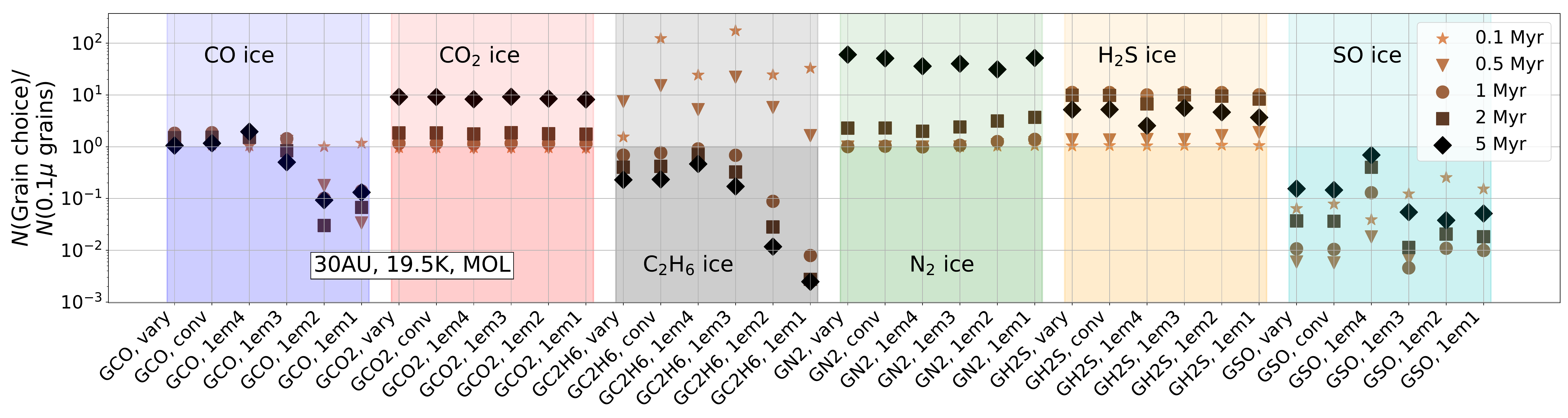}\\
    \caption{Similar to Fig. \ref{1_5_mol}, but for 30AU, and starting with molecular initial abundances. The converged (``conv'') grain size at 30AU is 3.85$\mu$m (see Table \ref{grain_sizes}). Please refer to caption of Fig. \ref{1_5_mol} for detailed description.}
    \label{30_mol}
\end{figure*}

\begin{figure*}
    \centering
    \includegraphics[width=.6\textwidth]{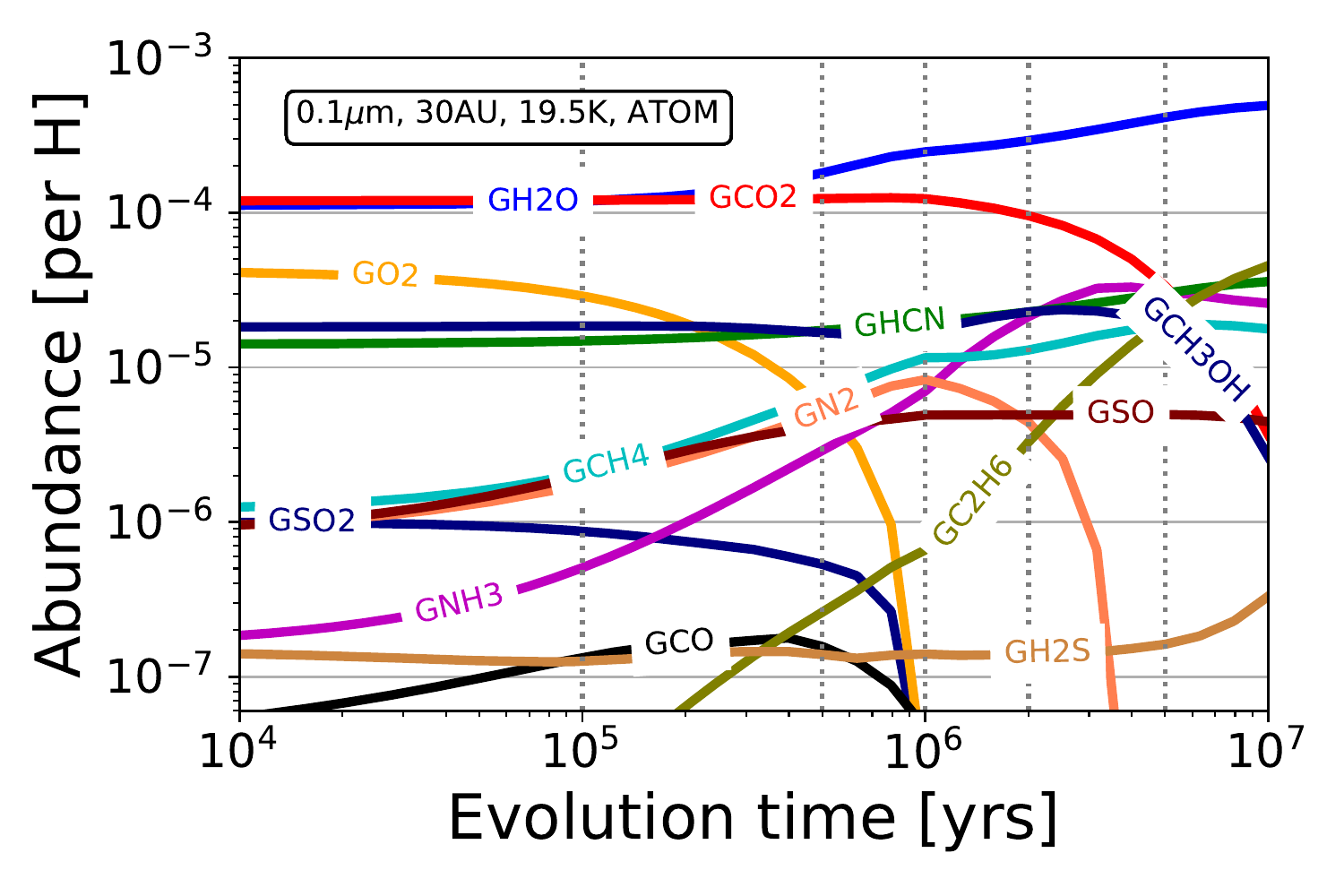}\\
    \includegraphics[width=1\textwidth]{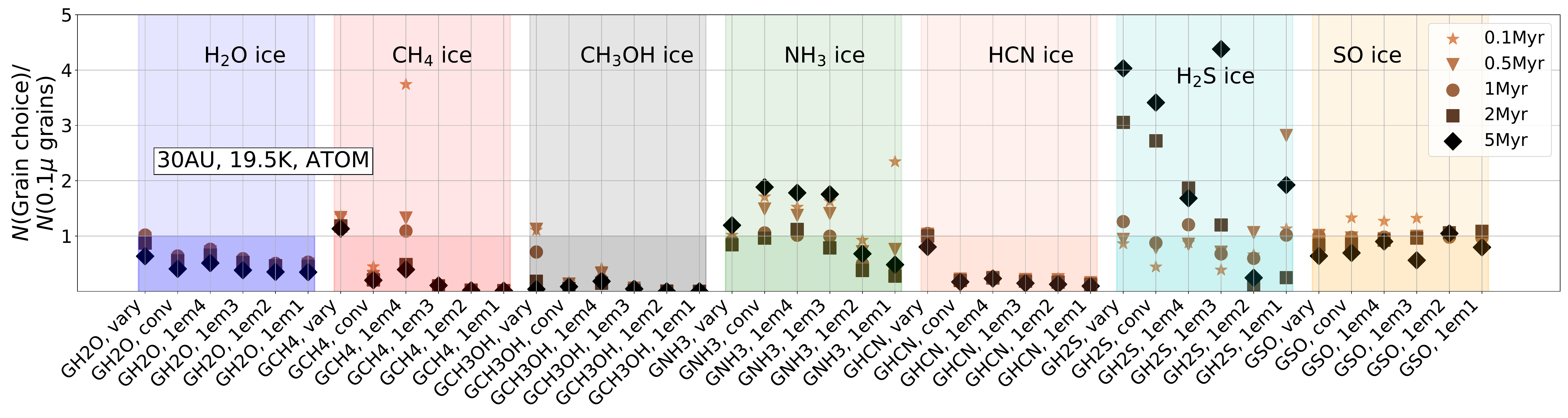}\\
    \includegraphics[width=1\textwidth]{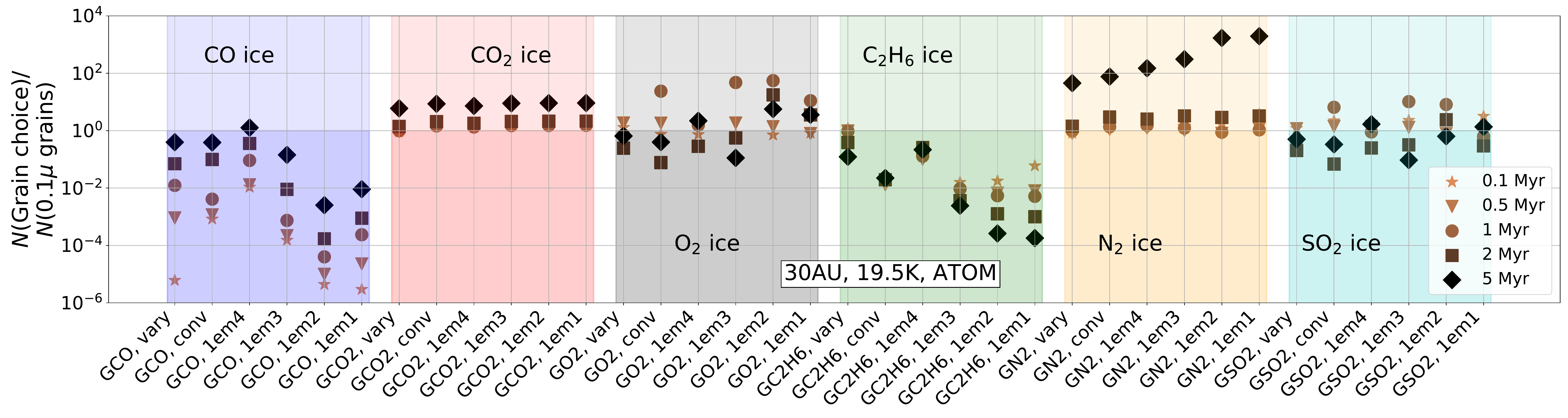}\\
    \caption{Similar to Fig. \ref{1_5_mol}, but for 30AU, and starting with atomic initial abundances. The converged (``conv'') grain size at 30AU is 3.85$\mu$m (see Table \ref{grain_sizes}). Please refer to caption of Fig. \ref{1_5_mol} for detailed description.}
    \label{30_atom}
\end{figure*}

Plots of absolute abundance evolution (abundance of molecule relative to abundance of hydrogen nuclei (H$_{\mathrm{nuc}}$) as a function of time for the dominant volatile carriers of elemental C, O, N and S are seen in Figs. \ref{1_5_mol} to \ref{30_atom}. Each figure focused on one combination of a disk midplane radius (1.5, 5, 20 or 30 AU) and an assumption for initial chemical abundances: ``MOL'' representing molecular initial abundances, and ``ATOM'' representing atomic initial abundances.

The top panel in each figure shows abundances as a function of evolution time for the dominant volatile carriers of C, O, N and S and spherical dust grains with a constant size of $R=0.1\mu$m. Note the logarithmic axes. Abundance profiles are dashed if a species is in the gas-phase for the given physical conditions, and solid if the species is in the ice-phase, as a part of the solid material. In the info boxes is also indicated the midplane temperatures of gas and grains at the given radius. The top plots for radii 5 and 30 AU, for molecular initial abundances (Figs. \ref{5_mol} and \ref{30_mol}), are models run with almost identical physical conditions and chemical assumptions to the models with results shown in Fig. 7, panels b and d in \citet{eistrup2018}. In the current study, however, three assumptions are different: most importantly, we have updated the molecular binding energy for atomic oxygen from $E_{\mathrm{bin}}$=800K to 1660K, in accordance with the findings of \citet{penteado2017}. Furthermore, we here assume a static physical disk environment throughout chemical evolution, as opposed to the physically evolving disk that was assumed in \citet{eistrup2018}, and we also now utilise a different time step resolution for the chemical modelling.


Our time steps here are adopted from the grain evolution models shown in Fig. \ref{grain_evol}. Because the grain evolution models capture the grain coagulation and fragmentation effects on short times, which are, in turn, shorter in the inner disk than in the outer, the time resolution is larger in the inner disk compared to the outer. At 1.5AU, a total of 345 time steps are utilised model chemical evolution from 1 to 10$^{7}$ year. At 30AU, only 179 time steps are used to model the same time scale. Furthermore, at 1.5AU, 265 time steps are used to chemically model the evolution from $\sim 300-3$k years, in order to capture the grain growth effects at this location. At 30AU, 97 time steps are used to model the evolution from $\sim$3k-30k years, which is the time scale for grain evolution at this location. The time step resolutions at 5AU and 2AU fall between these two described cases.

For comparison, Figure \ref{1mm} in the Appendix show the absolute abundance evolution at the four different midplane radii for a constant grain size of 1mm, thus the largest grain size option in this modelling framework.

The center and bottom panels of Figs. \ref{1_5_mol} to \ref{30_atom} each feature a number of color-shaded columns, with each color representing the evolution of one chemical species. Along the bottom of each of these two panels are categorical indicators: the name of the chemical species (matching that in the color-shaded region itself), and one of six categories:
\begin{itemize}
    \item ``vary'', representing chemical models run with evolving grain sizes.
    \item ``conv'', representing chemical models run with a constant grain size, where that size correspond to the converged value for that radius (see Table \ref{grain_sizes}).
    \item ``1em4'', representing chemical models run with a constant grain size of $R=10^{-4}$cm = 1$\mu$m.
    \item ``1em3'', representing chemical models run with a constant grain size of $R=10^{-3}$cm = 10$\mu$m.
    \item ``1em2'', representing chemical models run with a constant grain size of $R=10^{-2}$cm = 100$\mu$m.
    \item ``1em1'', representing chemical models run with a constant grain size of $R=10^{-1}$cm = 1mm.
\end{itemize}.
For each of these grain-size categories, each of the light-to-dark brown markers in the color-shaded regions represent the ratio between the abundance of that chemical species at at a given evolution time, \emph{assuming that choice of grain size}, and the abundance for the same species at the same evolution time \emph{assuming the nominal grain size of} 0.1$\mu$m. There are timestep markers for 0.1, 0.5, 1, 2 and 5 Myr (see legend insert), going from light to dark brown, and also changing marker shape for enhanced clarity (for reference, the five timesteps are indicated with vertical dotted lines in the tops panels). This choice of displaying the modelling results is intended to make clear what effect grain size choices has on each molecular species, at each radius, and on what timescales. For both center and bottom panels, the hue of the shading colors is lighter for $y$-axis ratios above unity, and darker for ratios below. This is to increase the ease of evaluating if a given grain size choice causes the of a certain molecule to be higher or lower than the abundance for the nominal grain size choice.

Lastly, the center panels feature a linear $y$-scale, whilst the scale is logarithmic in the bottom panels. This is due to the chemical species in the middle panels being most abundant overall, and therefore undergoing relatively smaller changes during evolution than is the case for the species in the bottom panels. The different scales are intended to enhance the dynamical ranges for the panels and ease their interpretation. The choice of which species to include in the middle and bottoms panels were made by eye by the main author for all eight figures.

\subsection{Chemical comparisons between 0.1 $\mu$m grains and other grain size choices}
In order to get a better impression of the effects that different grain size choices have on the abundance evolution for different chemical species, chemical abundances for different grain size choices relative to abundances for 0.1 $\mu$m-sized grains are show in Figures \ref{1_5_mol}-\ref{30_atom}. The chemical species considered are the dominant volatile carriers of carbon, oxygen, nitrogen and sulphur, and include: \ce{H2O}, \ce{CO2}, \ce{CH3OH}, CO, \ce{O2}, \ce{CH4}, \ce{C2H6}, HCN, \ce{NH3}, \ce{N2}, \ce{H2S}, \ce{SO2}, SO and \ce{S2}. This section will describe the key features of these figures.
\subsubsection{1.5AU, 120K, molecular initial abundances.}
\label{sec_5_mol}
The top panel of Fig. \ref{1_5_mol} features the time evolution for twelve different volatile species which, over a timescale of 10 Myr, are the main volatile carriers of C, N, O and S. The middle panel shows that \ce{H2O} ice is relatively more abundant with time and for all larger choices of grain sizes than the fiducial size of 0.1 micron. In fact, for all larger grain size choices (except for 1 micron) the \ce{H2O} ice abundance by 5 Myr is more than twice that for the fiducial case. Relating this to the top panel showing evolution, this corresponds to the \ce{H2O} abundance not decreasing after 2 Myr, but staying at roughly the initial abundances throughout evolution.

For CO gas in the middle panel, it is seen that its abundance for all grain size choices remains the same as for the fiducial size, for all time steps. Therefore, it is clear that the elemental oxygen needed for the increased water ice abundances at larger grain sizes stems from the remaining O-carrying species (\ce{CO2} gas, \ce{O2} gas, SO gas and \ce{SO2} ice), which are all lower for all grain size choice, relative to the 0.1 micron choice, for all evolution times. \ce{CH3OH} ice does feature some higher relative abundances if the grain sizes are varying, converged or 10 microns in size, by 2 Myr of evolution. However, for other sets of grain choices and evolution times, \ce{CH3OH} ice is also destroyed, contributing elemental O to increase the \ce{H2O} ice abundance. It is also seen in the top panel that the CO abundance is $>10^{-4}$ w.r.t H$_{\mathrm{nuc}}$, and given the global C/H ratio of 1.8$\cdot 10^{-4}$ w.r.t H$_{\mathrm{nuc}}$, it therefore follows that CO is the dominant carbon carrier.

Regarding nitrogen carriers, the abundances of \ce{N2} (middle panel) is largely unchanged for different grain size choices and different evolution times, and is the dominant carrier for all choices of parameters. For HCN in the lower panel, however, abundances increases ranging two to four orders-of-magnitude are seen by longer evolution times (2-5Myr) for all grain size choices, compared with the fiducial size. The source of the elemental nitrogen needed for this increase is seen to come mainly from \ce{NH3} gas, but also from destroyed \ce{N2}. Despite this increase in HCN abundance, it remains a minor carrier of nitrogen, as the baseline HCN abundance for the fiducial grain size after 2-5 Myr evolution is $<10^{-9}$, placing it outside of the dynamical range of the top panel.

For sulphur, it is clear that for grain sizes different from the fiducial size, sulphur is processed more into SO, and less into \ce{SO2} compared to the situation in the top panel. These two carriers feature similar abundances for all choices of grain size and evolution time. \ce{H2O} becomes an even smaller sulphur carrier, that what is shown in the top panel, compared with SO and \ce{SO2}. The exception for this is a constant grain size choice of 1 micron, for which the abundances of all three carriers are largely similar to that shown in the top panel. So only for grain sizes larger than 1 micron does \ce{SO2} and SO become the dominant sulphur carriers, and \ce{H2S} negligible. \ce{GSO2} and SO dominating the sulphur budget means that they collectively account for $\sim 1-2\%$ of elemental oxygen.

\subsubsection{1.5AU, 120K, atomic initial abundances.}

Starting from atomic initial abundances, the effects of grain growth at 1.5 AU are seen in Fig. \ref{1_5_atom}. Focusing first on \ce{H2O} ice (middle panel), it is seen that for all grain size choices, except 1mm (see top-right panel in Fig. \ref{1mm}) the abundances increasing over time by up to a factor of 1.7, compared with the fiducial grain size case. CO gas, \ce{N2} gas, \ce{SO2} ice and SO gas remain at the same abundance levels for all grain size choices and for all time steps, as were the cases for the fiducial grain size. This means that the the elemental oxygen contributing to the increase in \ce{H2O} ice (for all grain sizes except 1mm) is sourced from oxygen that would have otherwise been carried in \ce{CO2} and \ce{O2} gas. Both \ce{H2S} and HCN gas (lower panel) are lower in abundance for all grain size choices and timesteps than for the fiducial grain size choice, which means (comparing the bottom panel to the top panel) that both species are negligible in abundance for grain choices different from the fiducial size.

\subsubsection{5AU, 57K, molecular initial abundances.}

Slightly further out in the disk midplane, at 5AU and at a gas and grain temperature of 57K, the results of grain size choices of chemical evolution starting with a molecular composition can be seen in Fig. \ref{5_mol}. In the middle planes, it is clear that for all grain size choices other than the fiducial, there is an increase in \ce{H2O} ice abundances as a function of time. \ce{H2O} ice becomes up to 2.5 times more abundant by 5 Myr of evolution with grain sizes larger than the fiducial, which is a similar picture to what was seen for water in Section \ref{sec_5_mol} (Fig. \ref{1_5_mol}), namely that \ce{H2O} ice remains at approximately the same abundance throughtout evolution. Also alike the situation at 1.5AU, at 5AU this increase in abundance of \ce{H2O} ice is at the expense of the abundances of \ce{CO2} ice and \ce{O2} gas, with \ce{CO2} at 5AU being in the ice phase, where is was in the gas at 1.5AU.

CO gas generally is higher in abundance with larger grains sizes, except by 0.1 Myr. For 0.1 and 1mm sized grains, the CO gas abundances from 0.5 Myr and onwards is an order of magnitude higher than for the fiducial grain size at similar time steps. The \ce{CH4} gas abundance is also higher for larger grain sizes, compared to the fiducial size, for all but the 0.5 Myr timestep. For \ce{C2H6} ice, only a constant grain size of 1 micron results in higher abundances over all timesteps than the fiducial grain size choice.

For nitrogen carriers, most grain size choices, except 1mm, leads to higher abundances of \ce{NH3} and HCN ices at the expense of \ce{N2} gas production. In fact, it is seen that for smaller grain size choices, HCN ices becomes up to two orders of magnitude more abundant at late timesteps (2-5 Myr) than is the case for the fiducial grain size. Comparing this to the top figure, it shows that HCN ice reaches an abundance comporable to those of \ce{NH3} and \ce{N2} at late timestep.

For sulphur species, it is noticable that the abundance of \ce{H2S} gas is lower for most smaller grain size choices, compared to the fiducial grain size. Only for 0.1 and 1mm grain sizes does the \ce{H2S} abundance reach higher abundance than in the fiducial case, for timesteps 0.5-2Myr. It is seen on the abundance of \ce{SO2} ices that, except for the timestep at 0.1 Myr, the abundances for most grain size choice (except 1 micron), are higher at all timesteps 0.5 Myr and later, than the abundances for the fiducial grain size case.

\subsubsection{5AU, 57K, atomic initial abundances.}

Starting with atomic abundances at 57K with the fiducial 0.1 micron-sized grains, the top panel of Fig. \ref{5_atom} shows that \ce{CO2} ice and \ce{O2} gas quickly ($<10$ kyr) locks up the majority of C and O, and remains the main carriers of C and O throughout evolution. For larger grain size choices (except for ``vary''), it is seen that much of the O is processed into \ce{H2O} ($\sim10^{-4}$ ice, middle panel) at the expense of the \ce{CO2} ice throughout evolution (bottom panel), while \ce{O2} gas remains at $\pm$50\% of the level in the top panel, throughout evolution and for all choices of grain sizes. The C that is available for the decreasing abundance of \ce{CO2} ice for larger grain sizes is primarily processed into CO gas (bottom panel). Ultimately, these changes correspond to \ce{O2} gas, CO gas and \ce{H2O} ice all having similar abundances, when assuming larger grain sizes.

For N, larger choices of grain sizes generally means a decrease in HCN ice abundance, and an increase in \ce{NH3} ice and \ce{N2} gas abundances (except for 1mm grains, for which there is no increase in \ce{NH3} ice). For S, most of the S carried in \ce{SO2} ice in the top panel is processed into \ce{H2S} gas, for larger grain size choices.

\subsubsection{20AU, 25K, molecular initial abundances.}
\label{20mol}
At 25K, CO and \ce{N2} are the only carriers of C, O and N that are mainly in the gas-phase. In Fig. \ref{20_mol}, the top panel shows that \ce{H2O} ice (at $\sim 3\times 10^{-4}$ wrt H, thereby accounting for $\sim60\%$ of the elemental O) and \ce{CO2} ice (at $\sim 10^{-4}$ wrt H) are the dominant carriers of C and O throughout the chemical evolution. For grain size choices different from the fiducial, the middle panel shows that \ce{H2O} ice experiences a slight increase ($\sim20\%$), and \ce{CO2} ice a slight decrease, but they both remain the dominant carriers of C and O across grain size choices and time scales. The CO gas abundance, meanwhile, varies dramatically across evolution timescales in the bottom panel, but the abundance changes for CO gas as a function of evolution time is similar across different choices of grain sizes. In other words, for larger grain sizes, the choice of grain size does not matter as much for its abundance, as does the evolution time. The C left over from the decreasing CO gas and \ce{CO2} abundances are processed into species such as \ce{CH4} ice and HCN ice, for the largest grain sizes of 0.1-1mm. This processing into HCN ice, in turn, leads to a decrease in the abundances of \ce{NH3} ice and \ce{N2} gas at late evolutionary times, as seen in the middle panel.

At most evolutionary times and for grain sizes of 10micron (``1em3'') or smaller, some of the C from the decreasing \ce{CO2} ice abundance adds to the \ce{C2H6} ice abundances. The \ce{CH3OH} ice abundance decreases more with longer evolution time, down to two orders-of-magnitude lower by 10 Myr of chemical evolution for all grain size choices, than it is for the fiducial grain size. Sulphur is processed from SO ice into \ce{H2S} ice for larger grain size choices, which leads to the difference in the abundances of the two species in the top panel being leveled out, so they each carry roughly similar amounts of elemental S, by evolutionary times longer than 1 Myr.

\subsubsection{20AU, 25K, atomic initial abundances.}

Fig. \ref{20_atom} shows the evolution when starting with atomic abundances at 25K. Alike the situation in Sec. \ref{20mol}, \ce{H2O} ice and \ce{CO2} ice are the dominant carriers of C and O throughout evolution, for the fiducial grain size, as seen in the top panel. Interestingly, for larger grain size choices in the middle and bottom panels, the abundance of \ce{H2O} ice decreases more and more with longer evolution time, whereas \ce{CO2} ice experiences an abundance increase for all larger grain size choices, with the abundance growing larger with evolution time. After 0.5Myr of chemical evolution, the abundances of these two ices are roughly equivalent (both at $\sim2\times10^{-4}$ wrt H) and stay so until 5Myr of evolution. For the scenarios with varying grain sizes (``vary'') and constant grain size at 1 micron (``1em4''), \ce{CO2} is more abundant (by 50-100\%) than \ce{H2O} ice from 0.5Myr of evolution and onward.

While the O providing the increase in the \ce{CO2} ice abundance is mainly from a smaller production of \ce{H2O} ice for larger grain size choices than the fiducial, it is seen in the middle and bottom panels that the C carried in the \ce{CO2} ice for the larger grain size choices is a result of lower production of CO gas, \ce{CH4} ice, \ce{CH3OH} ice, \ce{C2H6} ice and HCN ice. That is to say, for larger grain size choices than the fiducial, \ce{H2O} ice and \ce{CO2} ice become the, by far, dominant carriers of C and O, accounting for around 99\% of both elements.

For varying grain sizes in the middle panel, N is processed more into \ce{NH3} ice and HCN ice with a lower production of \ce{N2} gas than in the fiducial case, but for all other larger grain size choices, HCN ice is lower in abundance than the two other species, and especially \ce{N2} gas is experiencing an increase in abundance, with \ce{NH3} ice sees a relatively smaller increase. For grain size choices larger than 10 microns (``1em3''), \ce{N2} gas is the most abundance N-carrying species until $\sim2$Myr of evolution, after which point \ce{NH3} ice is the most abundant.

For sulphur, larger grain size choices leads to more elemental S being carried in \ce{H2S} ice as compared to the fiducial grain size choice. \ce{SO2} ice and SO ice are both lower in abundance for larger grain size choices at all evolutionary times, than they were for the fiducial choice. However, for the two largest grain size choices, SO ice remains the dominant S-carrier for evolution times $\>2$Myr.

It is also noted that the top panel shows that a ratio of \ce{O2}-to-\ce{H2O} ice of $\>1\%$, for evolutionary times shorter than 0.5Myr. Furthermore, the bottom panel indicates that for all larger grain size choices, the \ce{O2} ice abundance is similar to the abundance for the fiducial grain choice after 1 Myr, and by 0.5 Myr, an order-of-magnitude larger than the abundance for the fiducial grain size choice, when grains of 0.1 or 1mm are assumed (``1em2'' and ``1em1'', respectively). For the same evolutionary times and grain choices, it is seen in the middle panel that the \ce{H2O} ice abundance is 25-40$\%$ lower than for the fiducial grain size choice. This reflects the underlying abundance condition: from the start of chemical evolution, and up until between 0.1-0.5 Myr of evolution, the abundance ratio \ce{O2}/\ce{H2O} $\sim 1-25\%$, for grain choices of 0.1 or 1mm.

\subsubsection{30AU, 19.5K, molecular initial abundances.}

At 19.5K, all considered species are in the ice-phase. For the fiducial grain size in the top panel of Fig. \ref{30_mol}, \ce{H2O} ice is the dominant carrier of O, by far, throughout evolution (growing in abundance from $\sim3-5\times10^{-4}$). For other grain size choices, there the \ce{H2O} ice abundance is slightly lower ($\sim20-30\%$) than for the fiducial grain size, at later evolutionary times, but still accounting for more than $60\%$ of elemental O. CO ice is experiencing an increase in abundance if the grain growth assumption is either the varying, the converged or the 1 micron grain size, relative to the fiducial gain size. For grain sizes larger than those, CO ice is less abundant throughout chemical evolution than is the case for the fiducial grain size case, with the exception of early evolutionary times ($<0.1$Myr), at which point the CO ice abundance is the same as it was in the fiducial grain size case.

C-carrying species such as HCN ice and \ce{CH3OH} ice generally reach lower abundances for larger grain size choices, whereas the abundances of \ce{CH4} ice and \ce{CO2} reach higher abundances for larger grain size choices. \ce{C2H6} ice reaches higher abundances at early evolutionary times (up to 0.5 Myr) than in the fiducial case, but is lower in abundance for all larger grain size choices for evolutionary times of 1 Myr and longer, as compared to the fiducial case.

For N-carrying species, larger grain size choices lead to higher abundances of \ce{N2} ice with longer chemical evolution (see bottom panel), whereas \ce{NH3} ice and HCN ice are slightly less abundant for larger grain sizes than for the fiducial grain size choice (see middle panel). However, \ce{NH3} ice is the most abundant N-carrying species at all evolutionary times and for all larger grain size choices, though after 1 Myr of evolution, all three molecules feature abundances within a factor of four of each other, and all are $>10^{-6}$ wrt H. For sulphur, larger grain size choices leads to lower SO ice abundances and higher \ce{H2S} abundances, relative to the fiducial grain size choice, for all evolution times. For grain size choices of 10 microns (``1em3'') or larger, \ce{H2S} ice reamins the dominant carrier throughout chemical evolution. However, for grain size choices of either the varying, the converged or the 1 micron grain size, relative to the fiducial gain size, SO ice becomes the dominant S-carrier after $\sim1$ Myr of evolution.

\subsubsection{30AU, 19.5K, atomic initial abundances.}

Alike the situation starting with molecular abundances at 19.5K, an assumption of atomic initial abundances also leads to less \ce{H2O} ice, and more \ce{CO2} ice produced for longer evolution times and larger grains size choices, than is the case for the fiducial grain size choice. Furthermore, since \ce{H2O} ice and \ce{CO2} ice in the top panel of Fig. \ref{30_atom} show similar abundances for evolution times shorter than 0.5 Myr, this means that, for larger grain size choices, \ce{CO2} ice is the more abundant of the two species until evolution times of $\sim1$ Myr.

The production of \ce{CO2} ice for larger grain size choices means than \ce{CH3OH} ice, CO ice, \ce{CH4} ice, and \ce{C2H6} ice are much less abundant for the choices than they were for the fiducial choice. The exception to this
is \ce{CH4} ice, which, for varying grain sizes, as well as for 1 micron grain size (``1em4''), is more abundant than for the fiducial grain size choice. \ce{O2} ice shows great variation in abundance depending on grain size choice and evolution times. However, for evolution times of 0.1 and 0.5 Myr (which, in the top panel, is the evolution times with significant \ce{O2} ice abundances) the \ce{O2} ice abundance is similar to the abundance for the fiducial grain size.

For N-carrying species, HCN ice is much less abundant for larger grain sizes than for the fiducial grain choice, whereas both \ce{NH3} ice and \ce{N2} ice are more abundance for grain size choices of 10 microns (``1em3'') or below. For larger grain size choices than that, only \ce{N2} ice is more abundant throughout evolution, compared to the fiducial grain size. For S-carrying species, \ce{H2O} ice is generally growing in abundance (for longer evolution times), at the expense of SO ice.

\section{Discussion}\label{disc}

In this section, the overall trends seen in the model results are highlighted. It is also discussed, how, and under which conditions the modelling of chemistry alongside with grain growth can be be simplified, for the purpose of easy adoption into astrochemical modelling frameworks.

\subsection{Dominant carriers of carbon and oxygen at different radii}

A key question to answer when implementing grain growth into a chemical kinetics modelling framework as done here regards, how the changing grain sizes affect which molecules are the dominant carriers of the chemical elements.

\subsubsection{Molecular initial abundances}
For 1.5AU and 5AU radii, the \ce{H2O} ice abundances, which were decreasing by $>$2Myr of time evolution time in the (top) plots of Figs. \ref{1_5_mol} and \ref{5_mol}, are seen to not undergo significant decrease at the late evolution times, for larger grain size choices (as seen by net increases in \ce{H2O} ice abundance for all evolution times and all larger grain size choices in the middle and bottom panels of the to figures). At 20AU and 30AU, the \ce{H2O} ice abundance remains largely at its initial abundance for all larger grain size choices. This means that for all modelling setups starting with molecular initial abundances, the \ce{H2O} ice abundance largely remains at its initial level, and is therefore the dominant molecule, and, in turn, the dominant carrier of elemental oxygen for all choices of grain size. This is featured, for the largest grain size choice of 1mm, in the left-hand column in Fig. \ref{1mm} in the Appendix.

Elemental carbon, on the other hand, becomes distributed across several different species, for larger grain size choices. At 1.5AU, CO gas is the dominant carrier for all grain sizes, including the 0.1$\mu$m. At 5AU, on the other hand, \ce{CO2} ice is the primary carrier, and \ce{C2H6} ice is the secondary carrier, after 1Myr of evolution, for a grain size of 1$\mu$m (``1em4''), but when assuming constant grain sizes of 100$\mu$m or 1mm, CO gas takes over as primary carrier of carbon, and \ce{CO2} ice becomes the secondary carrier. At 20AU, \ce{CO2} ice remains the primary carrier (by $>0.5$Myr of evolution) for grain sizes of 1$\mu$m-1mm, but the secondary carbon carrier for 1mm is \ce{C2H6} ice, whereas \ce{CH4} ice is the secondary carbon carrier for constant grain sizes of 100$\mu$m-1mm.

At 30AU, it is seen in the top panel of Fig. \ref{30_mol} that \ce{CO2} ice is the main carbon carrier until $\sim 4$Myr of chemical evolution, at which point \ce{C2H6} ice becomes the dominant carrier, with \ce{CH4} ice as the secondary carrier. For larger grain sizes of 100$\mu$m-1mm, \ce{CO2} ice is the dominant carrier until evolution times longer than 1Myr, after which point \ce{CH4} ice becomes the dominant carrier, with \ce{CO2} ice as secondary carrier.

In summary, while different grain size choices do changes the absolute abundance of all species, \ce{H2O} ice remains the main oxygen carrier. For carbon, however, larger grain size choices tend to change the main carbon carrier in the inner disk (1.5-5AU) from \ce{CO2} ice to CO gas, and from \ce{CO2} ice to \ce{CH4} ice in the outer, colder disk (20-30AU).

\subsubsection{Atomic initial abundances}

In the outer, colder disk midplane, at 20AU and 30AU, a comparison between Figs. \ref{20_atom}, \ref{30_atom} and \ref{1mm} shows that \ce{H2O} ice and \ce{CO2} ice are the dominant carriers of carbon and oxygen, both when assuming constant 0.1$\mu$m grains, and when assuming any larger grain size choices. However, while \ce{H2O} ice is the more abundant of the two for 0.1$\mu$m grains, the \ce{CO2} ice abundance increases to become the dominant carrier of both elemental carbon and oxygen for larger grain sizes.

For the warmest conditions, at 1.5AU, CO and \ce{O2} gas are the dominant carriers of elemental carbon and oxygen, for all choices of grain size. At 5AU, assuming 0.1$\mu$m size grains, \ce{CO2} ice and \ce{O2} gas are the dominant carriers, but for larger grain size choices, \ce{H2O} ice, \ce{CO2} ice, \ce{O2} gas and CO gas are all significant carriers of both elements, as seen in the bottom-right panel in Fig. \ref{1mm}, where all these molecules feature abundances without a factor of two of each other, for evolution times $<1$Myr. The temperature of 57K at 5AU, when assuming a purely atomic starting compositions, therefore might be a sweet spot for developing ice compositions of solid bodies that are not dominated by one, or even two species.

\subsection{Chemical evolution with evolving grain sizes v constant grain sizes}

This paper has explored and tested a range of different options for implementing grain growth into chemical kinetics in a protoplanetary disk midplane context. One important aspect to evaluate about the results is the extent to which accounting for an evolving grain population with an time-dependent grain-size distribution has significant impact on the chemical evolution, as compared with a constant grain population (in particular, a constant total grain-surface area), inspired by models of grain growth. In other words: can realistic chemical evolution be modelled based on a constant grain size (say, the converged grain size $R^A$ from the approach taken here) that is adopted from grain growth models, or is it necessary to evolve grain sizes ($R^A$) for each time step taken in a chemical kinetics model?

Before addressing this, it is noted that the grain size evolution at 1.5AU and 5AU reach the converged value for $R^A$ by 2.72kyr and 16.5kyr of evolution, respectively (see Table \ref{grain_sizes}), thus on a shorter timescale than 0.1Myr, which is the first evolution to for which the abundances of molecules are shown in the middle and bottom comparison panels in Figs. \ref{1_5_mol}-\ref{30_atom}. That is to say that the difference in chemical evolution between the setups with grain growth (``vary'') and the setup with a constant converged grain size (``conv'') for radii 1.5U and 5AU are only different from each other in the first thousands of years of chemical evolution, whereas from $\sim$20kyr-0.1Myr of evolution, the two cases are identical. This explains why the evolution of all considered species (for molecular initial abundances), are very similar across the ``vary'' and the ``conv'' cases, for 1.5AU and 5AU.

At 20AU and 30AU, the converged grain size is only reached by 0.13Myr and 0.39Myr respectively, which means that by 0.1Myr of evolution, the chemistry between the ``vary'' and ``conv''-cases have evolved with different grain sizes $R^A$ throughout. By 0.5Myr of chemical evolution at these two radii, the resulting abundances are more alike across the two grain size cases, and by 1Myr and longer, the abundances are largely similar.

In Table \ref{varyconv}, it is quantified how much the abundances of the three dominant carriers of carbon and oxygen in the ``conv'' grain-size case vary relative to their abundances in the ``vary'' grain-size cases. It is seen in this table that for all radii, and all evolution times, the variation between assuming evolving grain sizes and assuming a constant, converged grain size is $<1\%$. This means that, assuming that the timescale for grain growth to reach a converged representative grain size $R^A$ is less than two times the first chemical timescale considered (convergence of $R^A$ at 30AU is reached by $\sim$200kyr (see solid profiles in Fig. \ref{grain_evol}), and the first chemical timescale is 100kyr), then an assumption of the converged grain size throughout evolution will lead to the same chemical abundances of the main carriers of carbon and oxygen, as will grain sizes that evolve with time. For the purpose of modelling chemical evolution in protoplanetary disk midplane settings, it may therefore be an advantageous simplification to run chemical kinetics assuming the $R^A(t=\mathrm{final})$ of the grain population at the end of grain growth as a constant grain size, rather than accounting for each change in $R^A(t)$ that actually takes place for each chemical time step.

For the larger constant grain sizes considered here other other than ``vary'' and ``conv'', different evolution of abundances of all species is seen. It is the hope that these results and insights can serve as input for future research into chemical evolution, where larger grain sizes are considered.

\begin{table*}[]
    \centering
    \resizebox{\textwidth}{!}{%
    \begin{tabular}{l|ccc|ccc|ccc|ccc}
    & &1.5AU&&&5AU&&&20AU&&&30AU&\\
    \hline
    &&&&&&&&&&&&\\
    &0.1Myr&0.5Myr&1Myr&0.1Myr&0.5Myr&1Myr&0.1Myr&0.5Myr&1Myr&0.1Myr&0.5Myr&1Myr\\
    \hline
    &&&&&&&&&&&&\\
         \ce{H2O} ice&0.03\%&0\%&-0.06\%&-0.03\%&-0.03\%&-0.03\%&0.3\%&0.1\%&0.03\%& 0.07\%&0\%&-0.03\%\\
         \ce{CO2} ice&-&-&-&-&-&-&-1\%&-0.6\%& -0.2\%&0.3\%&-0.2\%&-0.05\% \\
         \ce{CO} gas&1\%&0.3\%&0.4\%&-0.4\%&-0.5\%&-0.7\%&-&-&-&-&-&- \\
    \end{tabular}}
    \caption{Variations in modelled abundances for when assuming the converged grain size as a constant size (``conv'') versus assuming the grain size to change for each chemical evolution timestep (``vary'') for the dominant carriers of C and O at the four radii. Calculated as $\frac{abun(conv)-abun(vary)}{abun(vary)}\times 100\%$, thus positive percentages mean that the abundance when using the converged grain size was larger than when using the varying grain size, and vice versa for negative percentages. For models run with molecular initial abundances. ``-'' indicates that a given phase of a species is irrelevant at a given radius, because it is either inside or outside the iceline of that species.}
    \label{varyconv}

\end{table*}

\subsection{Effects of grain growth on C/O ratios in the gas-phase}
An important factor when linking planet-formation in disks to observed exoplanets is the elemental ratios measured in disks and in exoplanet atmospheres. The C/O ratio has long been considered key to forging this link \citep[see, e.g.,][]{oberg2011co,madhu2014,eistrup2016,alidib2017,eistrup2018,molliere2022}, and recently similar ratios involving N and S have been suggested to overcome degenaracies resulting from the use of the C/O ratio, see \citet{turrini2021}. It is therefore interesting to consider how the different grain size choices in this work affects the C/O ratio.

\begin{figure*}
    \centering
    \includegraphics[width=.45\textwidth]{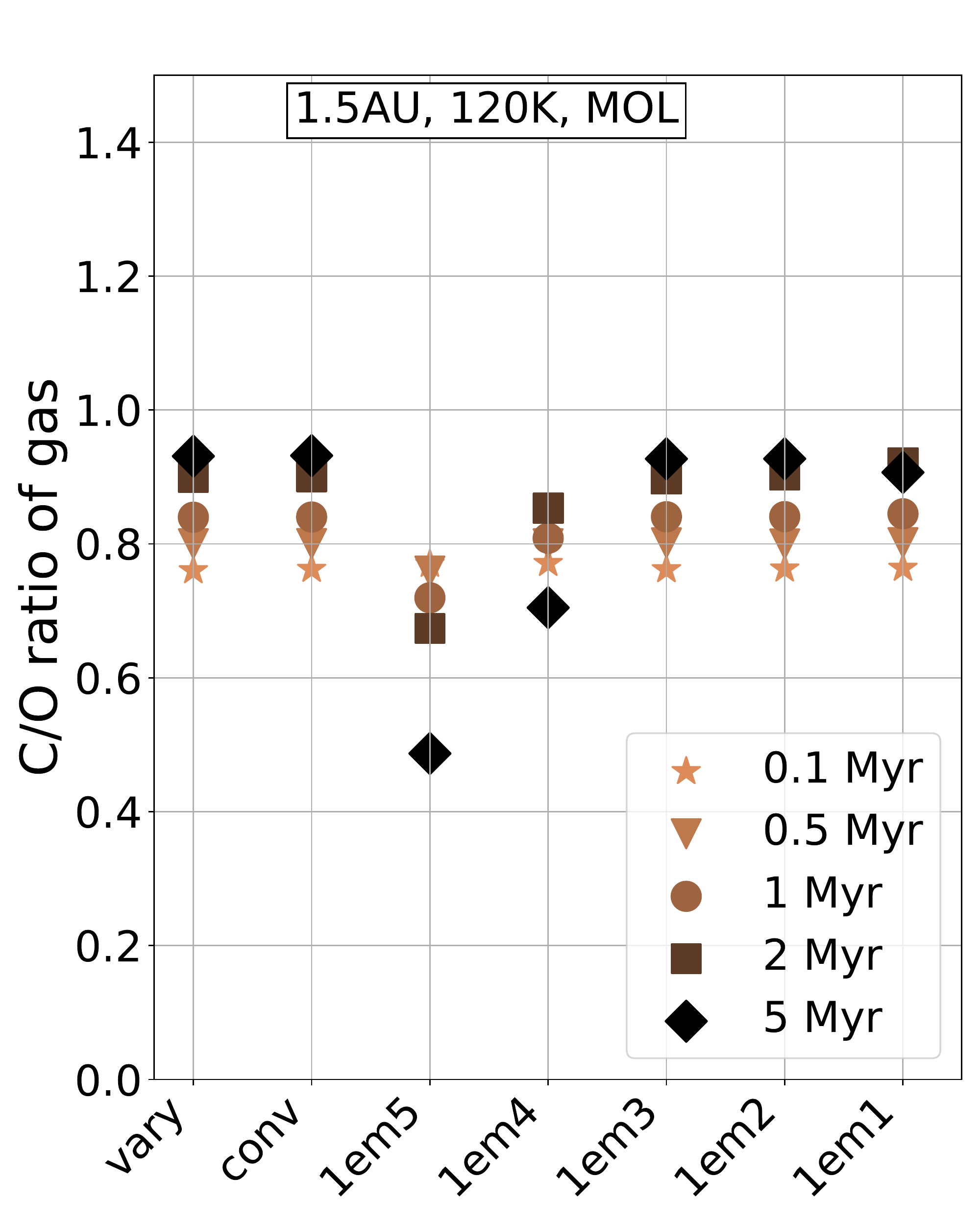}
    \includegraphics[width=.45\textwidth]{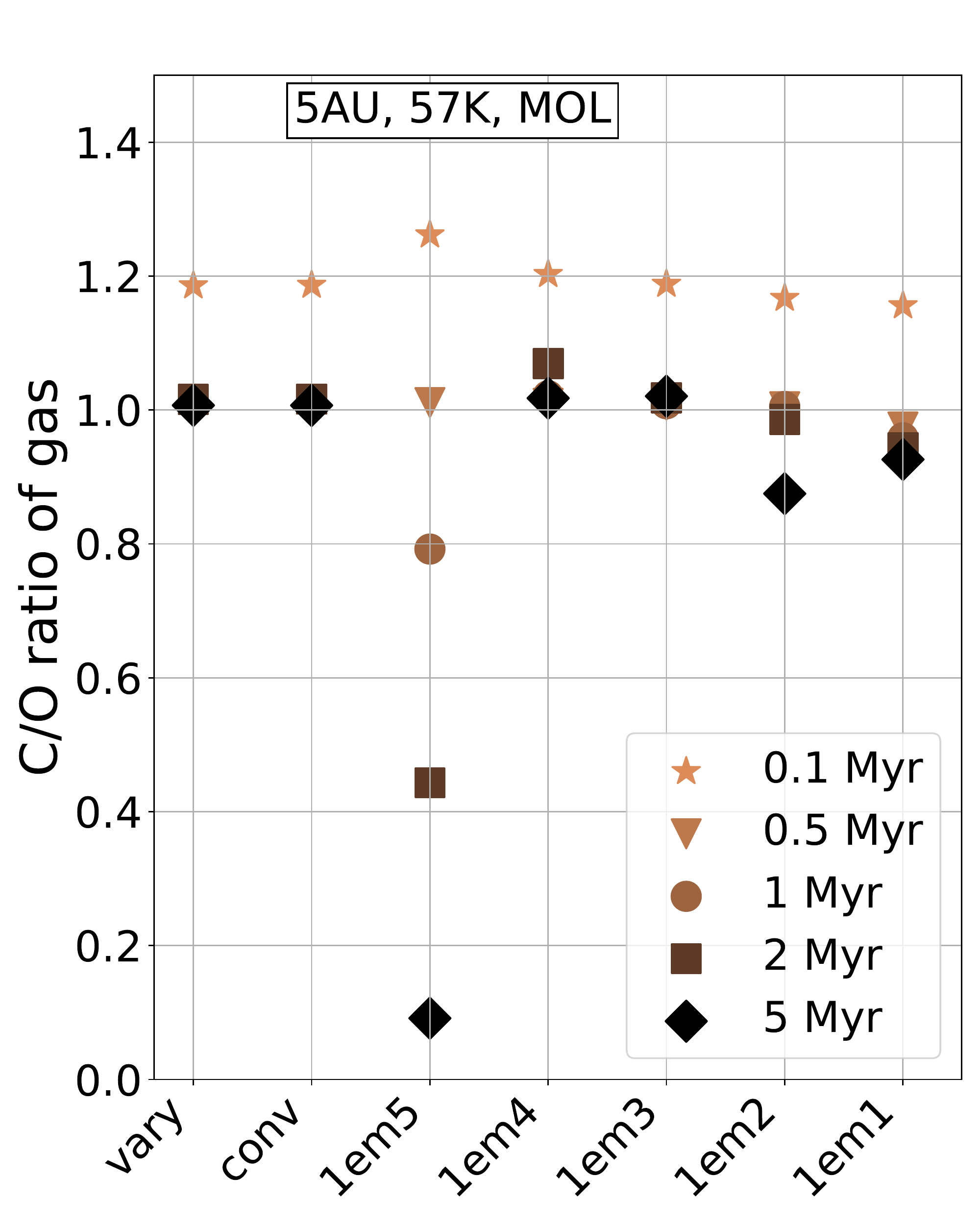}\\
    \caption{Carbon-to-oxygen ratios for gas-phase species, as a function of evolution time steps (colors and markers), for each of the seven grain choices (the third $x$-axis category, ``1em5'', is the fiducial grain size of 0.1$\mu$m). Left-hand panel is at 1.5AU, and right-hand panel is at 5AU. $y$-axis scales are identical in both panels. The models for 20AU and 30AU are not featured, because at 20AU only CO gas remains in the gas phase, and at 30AU all carbon and oxygen carriers are in the ice.}
    \label{c_to_o}
\end{figure*}

Fig. \ref{c_to_o} shows the C/O ratios for gas-phase species, with different colored markers indicating the evolution time. The seven types of grain sizes, for which chemical models were run, are indicated along the $x$-axis. The left-hand panel is for 1.5AU, and the right-hand panel is for 5AU. 20AU and 30AU are not relevant for gas-phase C/O ratios, as almost all carbon and oxygen-carrying species are frozen out at those temperatures (except for CO at 20AU). The third row in each panel shows the C/O ratio evolution for the fiducial grain size of 0.1$\mu$m. The evolution at both 1.5AU and 5AU for this grain size is comparable to the C/O ratio evolution plot inside and outside, respectively, of the \ce{CO2} iceline in Fig. 9.a in \citet{eistrup2018}. This is because the same chemical network and the same grain size was used, and only the physical structures vary slightly across these two studies.

It is clear that at both 1.5AU and 5AU, the C/O ratio evolution in the gas for the fiducial grain size is distinctly different than for the other grain size choices. At 1.5AU, the grain size choices ``vary'', ``conv'', ``1em3'', ``1em2'' and ``1em1'' all feature similar C/O ratio values with similar time dependencies. In all these case, the C/O ratio at 0.1Myr of evolution is similar to the cases for ``1em5'' and ``1em4'', but for longer evolution times, the C/O ratio value in the gas for the five larger grain size choices increases in time, from C/O$\sim$0.75 at 0.1Myr up to C/O$\sim$0.95 by 5Myr of evolution. This upward trend is different from the C/O ratio for the fiducial choice, where the time evolution trend goes from C/O$\sim$0.75 at 0.1Myr to C/O$\sim$0.5 at 5Myr. That means that the gas at 1.5AU becomes more carbon rich, relative to oxygen, for larger grain size choices. Because the total amounts of elemental C and O are constants, and the two elements can only exist as gas and as ice, this also means that the ice becomes more carbon poor, relative to oxygen. This goes opposite of the trend seen in Fig. 9.a of \citet{eistrup2018}.

At 5AU, the trend in C/O ratio value is similar (decreasing) for all grain size choices. However, for all other choices than the fiducial choice, the C/O ratio only decreases from C/O$\sim$1.2 to C/O$\sim$0.9-1.0. For the fiducial case, the final value is much lower, at C/O$\sim$0.1. Interestingly, at 5AU, little change to the C/O ratio values are seen over time, after 0.5Myr of evolution, for choices other than the fiducial choice.

Two effects influence these differences in the C/O ratio evolution for the fiducial grain size choice, and all other choices: First, for the fiducial grain size, elemental oxygen is processed from \ce{H2O} ice into \ce{O2} gas, which, due to the different molecular binding energies of these two species, serves to move elemental oxygen from the ice to the gas at midplane temperatures between 30-120K. In Figs. \ref{1_5_mol} and \ref{5_mol}, it is apparent that the decrease in \ce{H2O} ice abundance and the increase in \ce{O2} gas abundance which is seen in the top plots, are not seen for larger grain size choices. Second, for larger grain size choices, the effect of the ice phase reaction pathway
\begin{equation}
    \ce{H2O ->[\gamma_{\mathrm{CR}}] OH ->[CO] CO2}
\end{equation}
is reduced, whereby the CO gas abundance is generally higher, and the \ce{CO2} ice abundance generally lower, for larger grain size choices than for the fiducial choice. Along with the higher abundance of \ce{CH4} gas (at 5AU), this, all in all, maintains a large amount of elemental carbon (relative to oxygen) in the gas-phase, which keeps the C/O ratios in the gas relatively higher, compared with the case for the fiducial grain size choice. The reduced efficiency of the chemical reaction pathways described here can both be attributed to the fact that, for larger grain size choices, there are less overall grain surface area for ice chemistry to take place on, and thus the effect that grain surface-reactions and gas-grain interactions had for the fiducial case, is reduced for larger grain size choices.

\subsection{Comparison with previous work}

While modelling of chemical evolution in disks, and modelling of grain growth in disks, viewed apart, are active areas of research, there has been limited efforts attempting to combine the two. When done, the research goals across these efforts are not always aligned and, as such, the results from them can be difficult to compare with one another. \citet{vasyunin2011}, as well as some references therein, modelled chemical kinetics in a 2D protoplanetary disk, assuming two different constant grain population (and, hence, two different constant $R^A$) as well as a more realistic approach with grain evolution (grain sedimentation from upper disk layers to midplane, as well as grain growth). This work focused more on the effects of grain growth on resulting column densities of various species, by modelling chemical evolution in the upper layers of the disk, and not specifically the midplane. The results from this study therefore do not directly compare.

The work by \citet{gavino2021} used 12 different models for grains, where they varied the grains sizes, the grain temperatures, the UV radiation level, and the considered mechanisms for the formation of \ce{H2} on grain surfaces. Out of their 12 models, six assumed a constant grain size of 0.1$\mu$m, four took into account a full population of grains of different sizes where different sizes could be assigned different temperatures, and two used a single grain size representative of the grain-surface area in the full grain population, for three different radii. The latter two models, with a representative grain size $R^A$, follow the same concept as does the ``conv'' models in the current work. Whereas \citet{gavino2021} explores the effects of a full grain distribution with varying temperatures on the chemical evolution with no grain evolution, the current work explores the effects of a single grain size with a single temperature (per time step) with the single grain size evolving in time (but with no changing temperature).

It was predicted by \citet{gavino2021} that by including grain sizes larger than 0.1$\mu$m, carbon is processed into \ce{CO2} ice for warmer conditions, and into \ce{CH4} ice for colder conditions. This is in rough agreement with the trends seen in the current work, where larger grains lead to a change of major carbon carrier from \ce{CO2} ice to \ce{CH4} ice. \citet{gavino2021} also argues that a grain size distribution allowing for different temperatures for the different size populations allows for a production of COMs in the warmer grains that is not seen if a representative grain size $R^A$ is assumed with a constant temperature. In \citet{eistrup2018}, amongst others, it was also found that grains with temperatures in the ranges of 30K-40K produced more COMs than colder grains. However, it could be argued that assuming a representative grain size $R^A$ for each radius in the midplane (as is done in the current work), with more radial bins than here, and with dust temperatures changing from bin to bin could results in similar effects because bins with grain temperatures of 23K-40K would be encountered. This could facilitate the production of COMs, without assuming a grain size and temperature distribution at each radius.

\subsection{Impact of dust dynamics}\label{sec:dust_dynamics}
While relative velocities between aggregates were included in the dust coagulation routine described in Sect.~\ref{sec:dust_coagulation}, the models presented here are static in the sense that there is no material exchange from one region to another. As increasingly large aggregates (i.e. pebbles) are formed, however, their increased aerodynamical Stokes number causes them to partially decouple from the gas - settling vertically to the disk midplane and migrating radially inwards in the case of a smooth gas disk \citep[e.g.,][]{birnstiel2016, misener2019}. As most of the solid mass in contained in the larger aggregates, these processes can displace large amounts species present as ices, leading to complex and time-dependent redistribution of particularly those species that have their snowlines are crossed by drifting or settling pebbles \citep[e.g.,][]{cuzzi2004, krijt2018, boothilee2019}.

Recently, a treatment of vertical settling of dust grains from the upper layers of the disk towards the disk midplane was implemented into a chemical kinetics model by \citet{clepper2022graingrowth}. They modelled chemical evolution of ices on grains and pebbles in a midplane, accounting for the effect of turbulent mixing with the upper layers of the disk. Their grain growth model for the midplane included the continual inflow of grains that settled from the upper layers of the disk, as well as radial diffusion in the midplane. Neither of these effects are considered in this paper. However, they only considered one growth model with grains larger that 0.1$\mu$m, whereas this paper considers both one dynamic (``vary''), and several different larger constant grain sizes. \citet{clepper2022graingrowth} found CO gas to be depleted from the surface layers of a disk by 1-2 orders of magnitude, assuming a turbulence of  $\alpha=10^{-3}$. This is achieved by a combination of ice sequestration and decreasing UV opacity, both driven by pebble growth, they conclude. The radii they considered (30AU and 40AU), featured associated midplane temperatures of $\sim24$K and $\sim21$K, respectively (see their Fig.~2), which were similar to the temperatures considered here at 20AU and 30AU.

\citet{krijtbosman2020} used a compact set of chemical species (including \ce{H2O}, CO, \ce{CO2}, \ce{CH4} and \ce{CH3OH}) as primary carriers of elemental carbon and oxygen, with a network of ionisation, dissociation and hydrogenation reactions to exchange carbon and oxygen between the above five main carriers. This network being smaller, it enabled an easier implementation of chemical evolution into a grain evolution model, which included both sedimentation of grains to the midplane, growth and radial drift of the solids. They argued that all grain evolution effects they included were necessary in order to reproduce the observed low abundances of CO gas in protoplanetary disks \citep[e.g.][]{zhang2017,zhang2019,zhang2020co}. For the chemical evolution scenarios presented here, the CO gas abundance is generally similar to, or higher, if larger grain sizes than the fiducial size are assumed (at 1.5AU, 5AU and 20AU), and shorter evolution times are considered ($<2$Myr), highlighting the potential of dynamical effects to alter the resulting behaviour.


All these modelling efforts indicate that the fields of astrochemistry and grain evolution in protoplanetary disks are aiding each other constructively. This merging of the fields to improve the understanding of planet formation is important and necessary: in the coming years ALMA will keep expanding our understanding of the chemical (gas) condition for planet formation in protoplanetary disks, and \emph{JWST} will soon start exploring the compositions of protoplanetary disk ices, and the make-up of exoplanet atmospheres, through numerous observation programs. The wealth of new data from these facilities is poised to test the modelling frameworks and predictions that are now published or underway, and to provide a great leap for the understanding of how planets and their atmospheres form.

\section{Caveats of model assumptions}
\label{caveats}

\subsection{Treatment of charge conservation with grain growth}

In this work, the conservation of charge with grain growth was chosen to be accounted for by varying the electron abundance for each chemical and grain growth time step. Other options, as mentioned in Section \ref{methods}, would be to adjust the abundances of either cations or anions, or to allow for multiply-charged grains. In the case of multiply-charged grains, it would be interesting to investigate how an approach alike the one prescribed in \citet{fujii2011} could improve the treatment. This is left for future investigations.

\subsection{Fractal grain growth and fluffy aggregates: is spherical grain growth realistic?}

Realistic grain growth is not a process of growing from one sphere to another, but rather of micron-sized (or smaller) grains colliding and sticking \citep[see e.g.][]{windmark2012}. It is also suggested from theory that grain growth can lead to formation of ``fluffy aggregates'' \citep[][]{kataoka2015}. These are highly fractal structures which, despite resulting from collisions and sticking between its constituent micron-sized grains, have not compactified into a spherical shape. Fluffy aggregates thus feature effective grain surface areas comparable to the sum of the original grain surfaces areas of its constituents. In such a case, grains growing as fluffy aggregates do not reduce their grain surface area available for surface reaction, as compared to the case with 0.1$\mu$m-sized grains.

Realistic grain growth, however, is likely in-between compact, spherical growth and growth as fluffy aggregates (until these compactify). If grains in a protoplanetary disk midplane are assumed have a representative size (spherical or not) or, say, $R^{A}=10\mu$m, then it is likely that the 10$\mu$m case in this current work is underestimating the available grain surface area available for ice reactions. However, the current work provides chemical evolution results for 1$\mu$m grains as well, and so it is possible that the actual chemical evolution scenario can be assessed by assuming abundances in-between the results for 1$\mu$m and 10$\mu$m.

\section{Conclusion}

Protoplanetary disks facilitate both grain growth and chemical reactions leading to chemical evolution. However, until recently, models of grain growth did not account for chemical reactions, and models of chemical evolution did not account for grain growth. This paper set out to improve this treatment, by implementing a treatment of grain growth into the \textsc{Walsh} chemical kinetics-code. Results from modelling of grain growth have been utilised for modelling chemical evolution, both as grain sizes varying with time, and assuming a representative grain size throughout the chemical evolution. Discrete, constant grain sizes of 0.1$\mu$m, 1$\mu$m, 10$\mu$m, 100$\mu$m and 1mm have also been assumed, and their effects on the chemical evolution have been assessed.

Some key finding are summarised below:

\begin{itemize}
    \item Local, spherical grain growth, with no gain or loss of material over time, acts to reduced the total area on the surfaces of grains that is available for ice reaction of the grain surfaces to take place on. This reduces the efficiency of chemical reactions in the ice, while gas-phase reactions proceeds unaffected. This also means that ice chemistry produces fewer new molecules that could desorb into the gas phase, and partake in the gas-phase chemistry.
    \item Starting from a set of molecular initial abundances, all choices for grain growth lead to sustained or increased abundances of \ce{H2O} ice, which is the main carrier of elemental oxygen. In the inner, warmer disk midplane larger grain sizes lead to CO gas becoming the dominant carrier of elemental carbon, instead of \ce{CO2} ice. For lower temperatures in the outer, colder disk midplane, larger choices of grain sizes lead to \ce{CH4} ice becoming the dominant carbon carrier.
    \item The magnitude of change to the C/O ratio in the gas phase is smaller when considering grain growth or large grains, than when considering 0.1$\mu$-sized grains.
    \item Generally, assuming that the grain growth timescale is similar to or shorter than the chemical timescale, choosing a more realistic grain growth setup with grain sizes varying with time produces almost identical resulting abundances as does a more simplified setup, where a constant grain size is used throughout chemical evolution (a size with associated total grain surface area that mimics the area in the actual, evolved grain distribution). A simplified grain size setup in chemical kinetics codes with one constant (but realistic) grain size may therefore be favored.
\end{itemize}

The authors thank the referee, Tommaso Grassi, for suggesting valuable comments and improvements. CE thanks the Virginia Initiative for Cosmic Origins Postdoctoral Fellowship Program at the University of Virginia for support. Research for this paper was supported by the European Research Council under the Horizon 2020 Framework Program via the ERC Advanced Grant Origins 83 24 28.

\bibliographystyle{aa} 
\bibliography{refs.bib}

\newcommand{\noop}[1]{}
\begin{thebibliography}{57}
\expandafter\ifx\csname natexlab\endcsname\relax\def\natexlab#1{#1}\fi

\bibitem[{{Ali-Dib}(2017)}]{alidib2017}
{Ali-Dib}, M. 2017, MNRAS, 467, 2845

\bibitem[{{Birnstiel} {et~al.}(2015){Birnstiel}, {Andrews}, {Pinilla}, \&
  {Kama}}]{birnstiel2015}
{Birnstiel}, T., {Andrews}, S.~M., {Pinilla}, P., \& {Kama}, M. 2015, \apjl,
  813, L14

\bibitem[{{Birnstiel} {et~al.}(2018){Birnstiel}, {Dullemond}, {Zhu}, {Andrews},
  {Bai}, {Wilner}, {Carpenter}, {Huang}, {Isella}, {Benisty}, {P{\'e}rez}, \&
  {Zhang}}]{birnstiel2018_dsharp}
{Birnstiel}, T., {Dullemond}, C.~P., {Zhu}, Z., {et~al.} 2018, \apjl, 869, L45

\bibitem[{{Birnstiel} {et~al.}(2016){Birnstiel}, {Fang}, \&
  {Johansen}}]{birnstiel2016}
{Birnstiel}, T., {Fang}, M., \& {Johansen}, A. 2016, \ssr, 205, 41

\bibitem[{{Birnstiel} {et~al.}(2012){Birnstiel}, {Klahr}, \&
  {Ercolano}}]{birnstiel2012}
{Birnstiel}, T., {Klahr}, H., \& {Ercolano}, B. 2012, \aap, 539, A148

\bibitem[{{Birnstiel} {et~al.}(2011){Birnstiel}, {Ormel}, \&
  {Dullemond}}]{Birnstiel2011}
{Birnstiel}, T., {Ormel}, C.~W., \& {Dullemond}, C.~P. 2011, \aap, 525, A11

\bibitem[{{Bitsch} \& {Battistini}(2020)}]{bitschbattistini2020}
{Bitsch}, B. \& {Battistini}, C. 2020, A\&A, 633, A10

\bibitem[{{Bitsch} {et~al.}(2019){Bitsch}, {Izidoro}, {Johansen}, {Raymond},
  {Morbidelli}, {Lambrechts}, \& {Jacobson}}]{bitsch2019pebblesgiants}
{Bitsch}, B., {Izidoro}, A., {Johansen}, A., {et~al.} 2019, A\&A, 623, A88

\bibitem[{{Bitsch} \& {Johansen}(2016)}]{bitsch2016}
{Bitsch}, B. \& {Johansen}, A. 2016, A\&A, 590, A101

\bibitem[{{Bitsch} {et~al.}(2015){Bitsch}, {Lambrechts}, \&
  {Johansen}}]{bitsch2015}
{Bitsch}, B., {Lambrechts}, M., \& {Johansen}, A. 2015, A\&A, 582, A112

\bibitem[{{Bitsch} {et~al.}(2020){Bitsch}, {Trifonov}, \&
  {Izidoro}}]{bitschtrifonov2020}
{Bitsch}, B., {Trifonov}, T., \& {Izidoro}, A. 2020, \aap, 643, A66

\bibitem[{{Booth} \& {Ilee}(2019)}]{boothilee2019}
{Booth}, R.~A. \& {Ilee}, J.~D. 2019, MNRAS, 487, 3998

\bibitem[{{Brauer} {et~al.}(2008){Brauer}, {Dullemond}, \&
  {Henning}}]{brauer2008}
{Brauer}, F., {Dullemond}, C.~P., \& {Henning}, T. 2008, A\&A, 480, 859

\bibitem[{Clepper {et~al.}(2022)Clepper, Bergner, Bosman, Bergin, \&
  Ciesla}]{clepper2022graingrowth}
Clepper, E.~V., Bergner, J.~B., Bosman, A.~D., Bergin, E., \& Ciesla, F.~J.
  2022 [\eprint{Arxiv:2202.00524v1}]

\bibitem[{{Cridland} {et~al.}(2016){Cridland}, {Pudritz}, \&
  {Alessi}}]{cridland2016}
{Cridland}, A.~J., {Pudritz}, R.~E., \& {Alessi}, M. 2016, MNRAS, 461, 3274

\bibitem[{{Cridland} {et~al.}(2017){Cridland}, {Pudritz}, {Birnstiel},
  {Cleeves}, \& {Bergin}}]{cridland2017}
{Cridland}, A.~J., {Pudritz}, R.~E., {Birnstiel}, T., {Cleeves}, L.~I., \&
  {Bergin}, E.~A. 2017, MNRAS, 469, 3910

\bibitem[{{Cuzzi} \& {Zahnle}(2004)}]{cuzzi2004}
{Cuzzi}, J.~N. \& {Zahnle}, K.~J. 2004, \apj, 614, 490

\bibitem[{{Draine} \& {Sutin}(1987)}]{drainesutin1987}
{Draine}, B.~T. \& {Sutin}, B. 1987, \apj, 320, 803

\bibitem[{{Dr{\k{a}}{\.z}kowska} {et~al.}(2014){Dr{\k{a}}{\.z}kowska},
  {Windmark}, \& {Dullemond}}]{drazkowska2014}
{Dr{\k{a}}{\.z}kowska}, J., {Windmark}, F., \& {Dullemond}, C.~P. 2014, \aap,
  567, A38

\bibitem[{{Drummond} {et~al.}(2019){Drummond}, {Carter}, {H{\'e}brard},
  {Mayne}, {Sing}, {Evans}, \& {Goyal}}]{drummond2019}
{Drummond}, B., {Carter}, A.~L., {H{\'e}brard}, E., {et~al.} 2019, MNRAS, 486,
  1123

\bibitem[{{Eistrup} {et~al.}(2016){Eistrup}, {Walsh}, \& {van
  Dishoeck}}]{eistrup2016}
{Eistrup}, C., {Walsh}, C., \& {van Dishoeck}, E.~F. 2016, \aap, 595, A83

\bibitem[{{Eistrup} {et~al.}(2018){Eistrup}, {Walsh}, \& {van
  Dishoeck}}]{eistrup2018}
{Eistrup}, C., {Walsh}, C., \& {van Dishoeck}, E.~F. 2018, \aap, 613, A14

\bibitem[{{Fogel} {et~al.}(2011){Fogel}, {Bethell}, {Bergin}, {Calvet}, \&
  {Semenov}}]{fogel2011}
{Fogel}, J. K.~J., {Bethell}, T.~J., {Bergin}, E.~A., {Calvet}, N., \&
  {Semenov}, D. 2011, \apj, 726, 29

\bibitem[{Fraser {et~al.}(2001)Fraser, Collings, McCoustra, \&
  Williams}]{fraser2001}
Fraser, H.~J., Collings, M.~P., McCoustra, M. R.~S., \& Williams, D.~A. 2001,
  Monthly Notices of the Royal Astronomical Society, 327, 1165–1172

\bibitem[{{Fujii} {et~al.}(2011){Fujii}, {Okuzumi}, \& {Inutsuka}}]{fujii2011}
{Fujii}, Y.~I., {Okuzumi}, S., \& {Inutsuka}, S.-i. 2011, ApJ, 743, 53

\bibitem[{{Gavino} {et~al.}(2021){Gavino}, {Dutrey}, {Wakelam}, {Guilloteau},
  {Kobus}, {Wolf}, {Iqbal}, {Di Folco}, {Chapillon}, \&
  {Pi{\'e}tu}}]{gavino2021}
{Gavino}, S., {Dutrey}, A., {Wakelam}, V., {et~al.} 2021, arXiv e-prints,
  arXiv:2106.05888

\bibitem[{{Grassi} {et~al.}(2014){Grassi}, {Bovino}, {Schleicher}, {Prieto},
  {Seifried}, {Simoncini}, \& {Gianturco}}]{grassi2014}
{Grassi}, T., {Bovino}, S., {Schleicher}, D.~R.~G., {et~al.} 2014, MNRAS, 439,
  2386

\bibitem[{{Gravity Collaboration} {et~al.}(2020){Gravity Collaboration},
  {Nowak}, {Lacour}, {Molli{\`e}re}, {Wang}, {Charnay}, {van Dishoeck},
  {Abuter}, {Amorim}, {Berger}, {Beust}, {Bonnefoy}, {Bonnet}, {Brandner},
  {Buron}, {Cantalloube}, {Collin}, {Chapron}, {Cl{\'e}net}, {Coud{\'e} Du
  Foresto}, {de Zeeuw}, {Dembet}, {Dexter}, {Duvert}, {Eckart}, {Eisenhauer},
  {F{\"o}rster Schreiber}, {F{\'e}dou}, {Garcia Lopez}, {Gao}, {Gendron},
  {Genzel}, {Gillessen}, {Hau{\ss}mann}, {Henning}, {Hippler}, {Hubert},
  {Jocou}, {Kervella}, {Lagrange}, {Lapeyr{\`e}re}, {Le Bouquin}, {L{\'e}na},
  {Maire}, {Ott}, {Paumard}, {Paladini}, {Perraut}, {Perrin}, {Pueyo}, {Pfuhl},
  {Rabien}, {Rau}, {Rodr{\'\i}guez-Coira}, {Rousset}, {Scheithauer},
  {Shangguan}, {Straub}, {Straubmeier}, {Sturm}, {Tacconi}, {Vincent},
  {Widmann}, {Wieprecht}, {Wiezorrek}, {Woillez}, {Yazici}, \&
  {Ziegler}}]{gravity2020betapic}
{Gravity Collaboration}, {Nowak}, M., {Lacour}, S., {et~al.} 2020, A\&A, 633,
  A110

\bibitem[{{Hasegawa} \& {Herbst}(1993)}]{hasegawa1993}
{Hasegawa}, T.~I. \& {Herbst}, E. 1993, MNRAS, 261, 83

\bibitem[{{Herbst}(2021)}]{herbst2021review}
{Herbst}, E. 2021, Frontiers in Astronomy and Space Sciences, 8, 207

\bibitem[{{Kataoka} {et~al.}(2015){Kataoka}, {Muto}, {Momose}, {Tsukagoshi},
  {Fukagawa}, {Shibai}, {Hanawa}, {Murakawa}, \& {Dullemond}}]{kataoka2015}
{Kataoka}, A., {Muto}, T., {Momose}, M., {et~al.} 2015, \apj, 809, 78

\bibitem[{{Krijt} {et~al.}(2020){Krijt}, {Bosman}, {Zhang}, {Schwarz},
  {Ciesla}, \& {Bergin}}]{krijtbosman2020}
{Krijt}, S., {Bosman}, A.~D., {Zhang}, K., {et~al.} 2020, ApJ, 899, 134

\bibitem[{{Krijt} {et~al.}(2016){Krijt}, {Ciesla}, \&
  {Bergin}}]{krijt2016_dustgrowth}
{Krijt}, S., {Ciesla}, F.~J., \& {Bergin}, E.~A. 2016, \apj, 833, 285

\bibitem[{{Krijt} {et~al.}(2018){Krijt}, {Schwarz}, {Bergin}, \&
  {Ciesla}}]{krijt2018}
{Krijt}, S., {Schwarz}, K.~R., {Bergin}, E.~A., \& {Ciesla}, F.~J. 2018, \apj,
  864, 78

\bibitem[{{Lambrechts} \& {Johansen}(2012)}]{lambrechts2012}
{Lambrechts}, M. \& {Johansen}, A. 2012, A\&A, 544, A32

\bibitem[{{Line} {et~al.}(2021){Line}, {Brogi}, {Bean}, {Gandhi}, {Zalesky},
  {Parmentier}, {Smith}, {Mace}, {Mansfield}, {Kempton}, {Fortney}, {Shkolnik},
  {Patience}, {Rauscher}, {D{\'e}sert}, \& {Wardenier}}]{line2021ctoo}
{Line}, M.~R., {Brogi}, M., {Bean}, J.~L., {et~al.} 2021, \nat, 598, 580

\bibitem[{{Madhusudhan} {et~al.}(2014){Madhusudhan}, {Amin}, \&
  {Kennedy}}]{madhu2014}
{Madhusudhan}, N., {Amin}, M.~A., \& {Kennedy}, G.~M. 2014, ApJl, 794, L12

\bibitem[{{Misener} {et~al.}(2019){Misener}, {Krijt}, \&
  {Ciesla}}]{misener2019}
{Misener}, W., {Krijt}, S., \& {Ciesla}, F.~J. 2019, \apj, 885, 118

\bibitem[{{Molli{\`e}re} {et~al.}(2022){Molli{\`e}re}, {Molyarova}, {Bitsch},
  {Henning}, {Schneider}, {Kreidberg}, {Eistrup}, {Burn}, {Nasedkin},
  {Semenov}, {Mordasini}, {Schlecker}, {Schwarz}, {Lacour}, {Nowak}, \&
  {Schulik}}]{molliere2022}
{Molli{\`e}re}, P., {Molyarova}, T., {Bitsch}, B., {et~al.} 2022, arXiv
  e-prints, arXiv:2204.13714

\bibitem[{{Molli{\`e}re} {et~al.}(2020){Molli{\`e}re}, {Stolker}, {Lacour},
  {Otten}, {Shangguan}, {Charnay}, {Molyarova}, {Nowak}, {Henning}, {Marleau},
  {Semenov}, {van Dishoeck}, {Eisenhauer}, {Garcia}, {Garcia Lopez}, {Girard},
  {Greenbaum}, {Hinkley}, {Kervella}, {Kreidberg}, {Maire}, {Nasedkin},
  {Pueyo}, {Snellen}, {Vigan}, {Wang}, {de Zeeuw}, \& {Zurlo}}]{molliere2020}
{Molli{\`e}re}, P., {Stolker}, T., {Lacour}, S., {et~al.} 2020, A\&A, 640, A131

\bibitem[{{Mordasini} {et~al.}(2016){Mordasini}, {van Boekel}, {Molli{\`e}re},
  {Henning}, \& {Benneke}}]{mordasini2016}
{Mordasini}, C., {van Boekel}, R., {Molli{\`e}re}, P., {Henning}, T., \&
  {Benneke}, B. 2016, ApJ, 832, 41

\bibitem[{{Notsu} {et~al.}(2020){Notsu}, {Eistrup}, {Walsh}, \&
  {Nomura}}]{notsu_eistrup2020}
{Notsu}, S., {Eistrup}, C., {Walsh}, C., \& {Nomura}, H. 2020, MNRAS
  [\eprint[arXiv]{2009.09444}]

\bibitem[{{{\"O}berg} {et~al.}(2011){{\"O}berg}, {Murray-Clay}, \&
  {Bergin}}]{oberg2011co}
{{\"O}berg}, K.~I., {Murray-Clay}, R., \& {Bergin}, E.~A. 2011, ApJl, 743, L16

\bibitem[{{Ormel} \& {Cuzzi}(2007)}]{ormelcuzzi2007}
{Ormel}, C.~W. \& {Cuzzi}, J.~N. 2007, \aap, 466, 413

\bibitem[{{Penteado} {et~al.}(2017){Penteado}, {Walsh}, \&
  {Cuppen}}]{penteado2017}
{Penteado}, E.~M., {Walsh}, C., \& {Cuppen}, H.~M. 2017, ApJ, 844, 71

\bibitem[{{Ruaud} {et~al.}(2016){Ruaud}, {Wakelam}, \& {Hersant}}]{ruaud2016}
{Ruaud}, M., {Wakelam}, V., \& {Hersant}, F. 2016, \mnras, 459, 3756

\bibitem[{{Testi} {et~al.}(2014){Testi}, {Birnstiel}, {Ricci}, {Andrews},
  {Blum}, {Carpenter}, {Dominik}, {Isella}, {Natta}, {Williams}, \&
  {Wilner}}]{testi2014}
{Testi}, L., {Birnstiel}, T., {Ricci}, L., {et~al.} 2014, Protostars and
  Planets VI, 339

\bibitem[{{Trapman} {et~al.}(2019){Trapman}, {Facchini}, {Hogerheijde}, {van
  Dishoeck}, \& {Bruderer}}]{trapman2019}
{Trapman}, L., {Facchini}, S., {Hogerheijde}, M.~R., {van Dishoeck}, E.~F., \&
  {Bruderer}, S. 2019, A\&A, 629, A79

\bibitem[{{Turrini} {et~al.}(2021){Turrini}, {Schisano}, {Fonte}, {Molinari},
  {Politi}, {Fedele}, {Pani{\'c}}, {Kama}, {Changeat}, \&
  {Tinetti}}]{turrini2021}
{Turrini}, D., {Schisano}, E., {Fonte}, S., {et~al.} 2021, ApJ, 909, 40

\bibitem[{{Vasyunin} {et~al.}(2011){Vasyunin}, {Wiebe}, {Birnstiel},
  {Zhukovska}, {Henning}, \& {Dullemond}}]{vasyunin2011}
{Vasyunin}, A.~I., {Wiebe}, D.~S., {Birnstiel}, T., {et~al.} 2011, \apj, 727,
  76

\bibitem[{{Walsh} {et~al.}(2015){Walsh}, {Nomura}, \& {van
  Dishoeck}}]{walsh2015}
{Walsh}, C., {Nomura}, H., \& {van Dishoeck}, E. 2015, \aap, 582, A88

\bibitem[{{Windmark} {et~al.}(2012){Windmark}, {Birnstiel}, {Ormel}, \&
  {Dullemond}}]{windmark2012}
{Windmark}, F., {Birnstiel}, T., {Ormel}, C.~W., \& {Dullemond}, C.~P. 2012, A,
  544, L16

\bibitem[{{Yu} {et~al.}(2017){Yu}, {Evans}, {Dodson-Robinson}, {Willacy}, \&
  {Turner}}]{yu2017}
{Yu}, M., {Evans}, II, N.~J., {Dodson-Robinson}, S.~E., {Willacy}, K., \&
  {Turner}, N.~J. 2017, ApJ, 841, 39

\bibitem[{{Zhang} {et~al.}(2017){Zhang}, {Bergin}, {Blake}, {Cleeves}, \&
  {Schwarz}}]{zhang2017}
{Zhang}, K., {Bergin}, E.~A., {Blake}, G.~A., {Cleeves}, L.~I., \& {Schwarz},
  K.~R. 2017, Nature Astronomy, 1, 0130

\bibitem[{{Zhang} {et~al.}(2019){Zhang}, {Bergin}, {Schwarz}, {Krijt}, \&
  {Ciesla}}]{zhang2019}
{Zhang}, K., {Bergin}, E.~A., {Schwarz}, K., {Krijt}, S., \& {Ciesla}, F. 2019,
  \apj, 883, 98

\bibitem[{Zhang {et~al.}(2020)Zhang, Schwarz, \& Bergin}]{zhang2020co}
Zhang, K., Schwarz, K.~R., \& Bergin, E.~A. 2020, The Astrophysical Journal,
  891, L17

\bibitem[{{Zsom} \& {Dullemond}(2008)}]{zsom2008}
{Zsom}, A. \& {Dullemond}, C.~P. 2008, \aap, 489, 931

\end{thebibliography}

\begin{appendix}

\section{Chemical abundances as a function of time, for a constant grain size $R_\mathrm{grain}=1$mm.}

\begin{figure*}
    \centering
    \includegraphics[width=.48\textwidth]{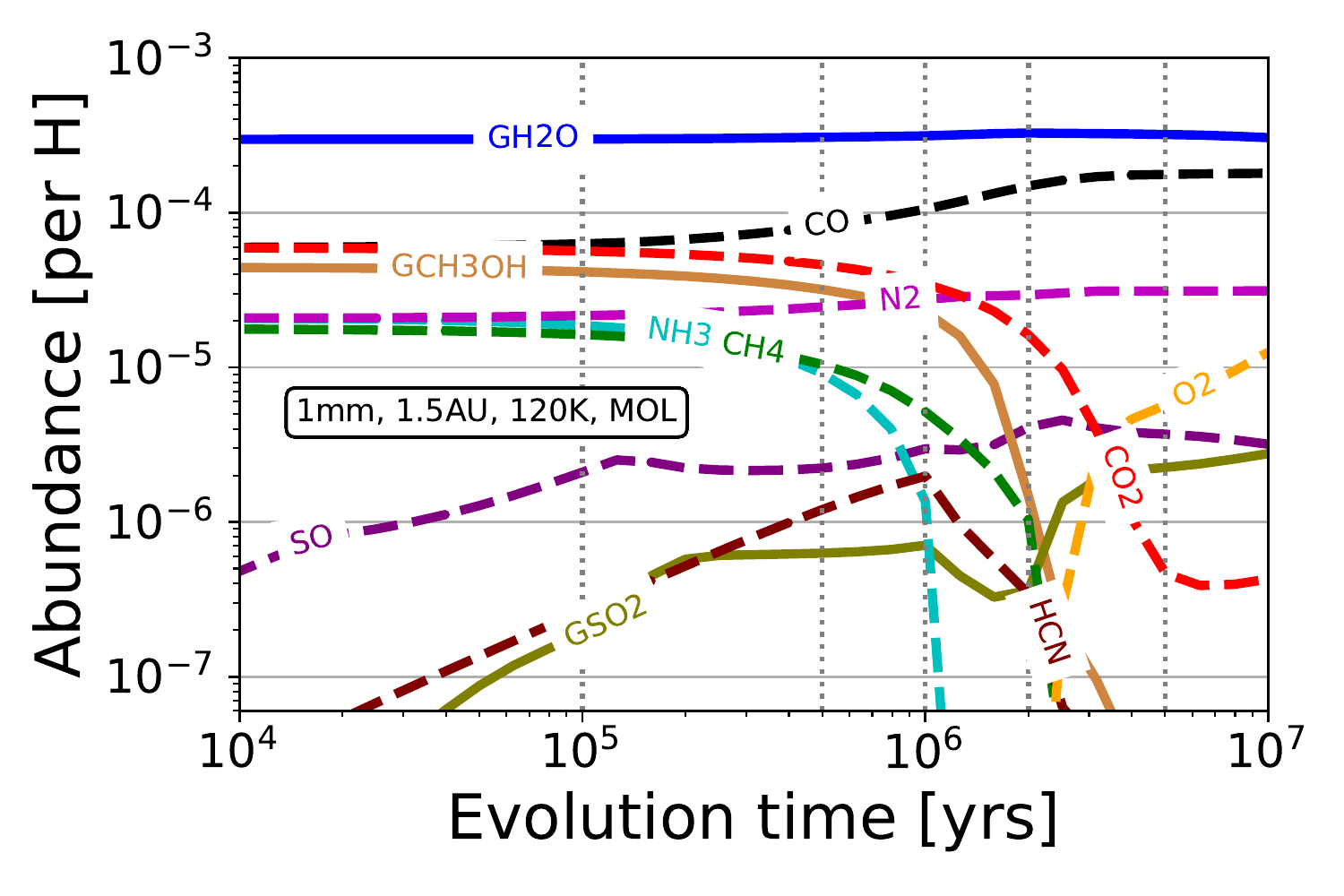}
    \includegraphics[width=.48\textwidth]{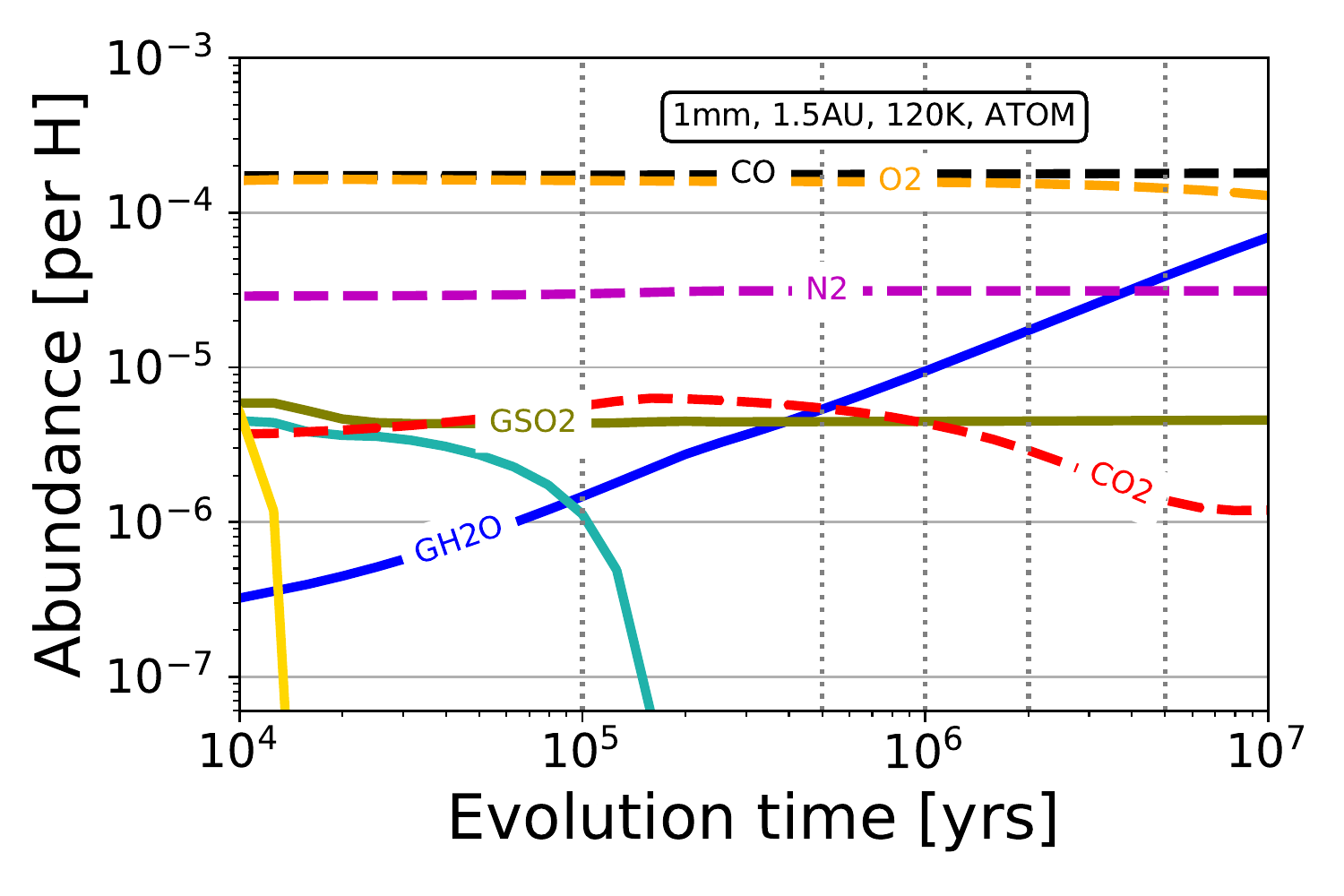}\\
    \includegraphics[width=.48\textwidth]{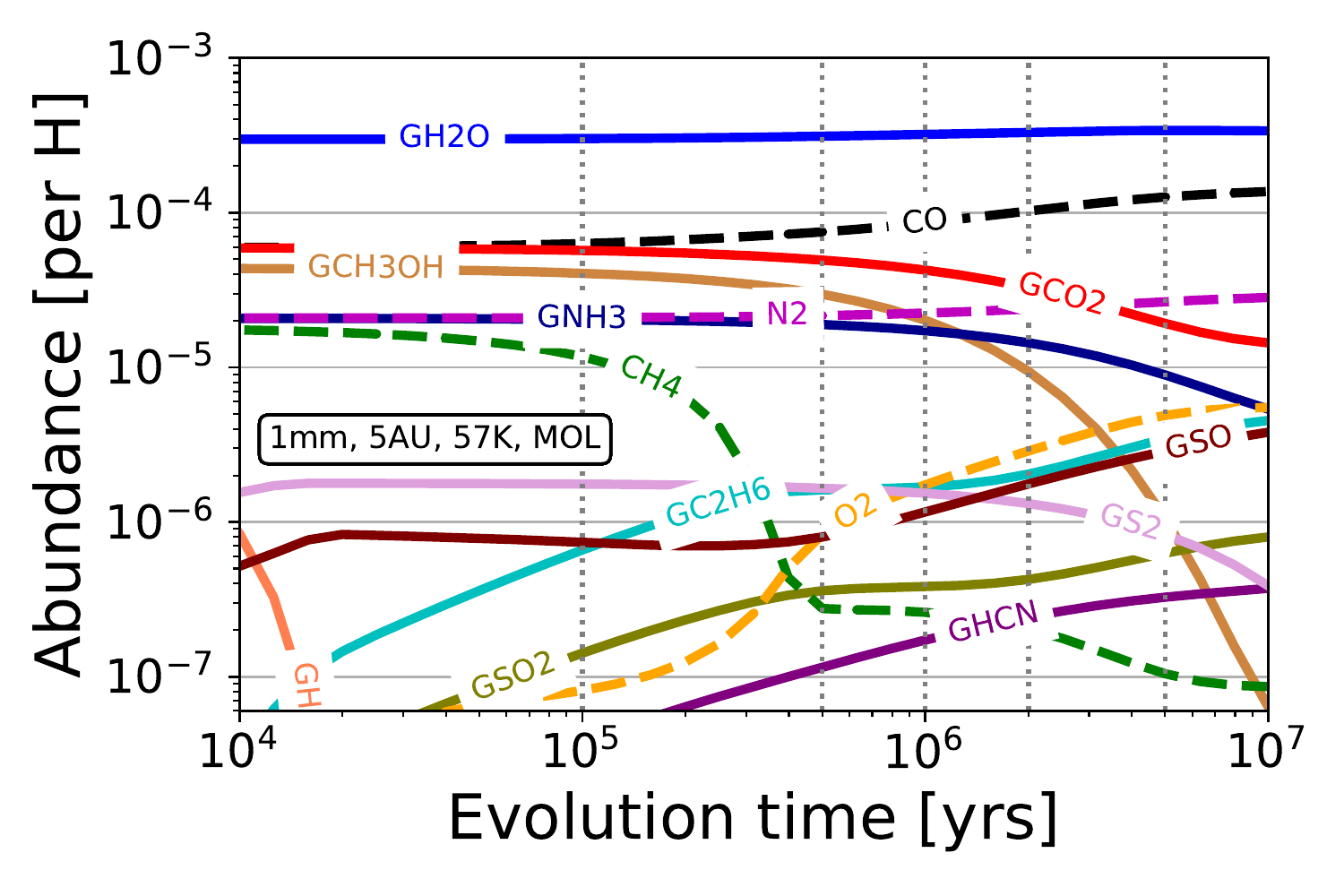}
    \includegraphics[width=.48\textwidth]{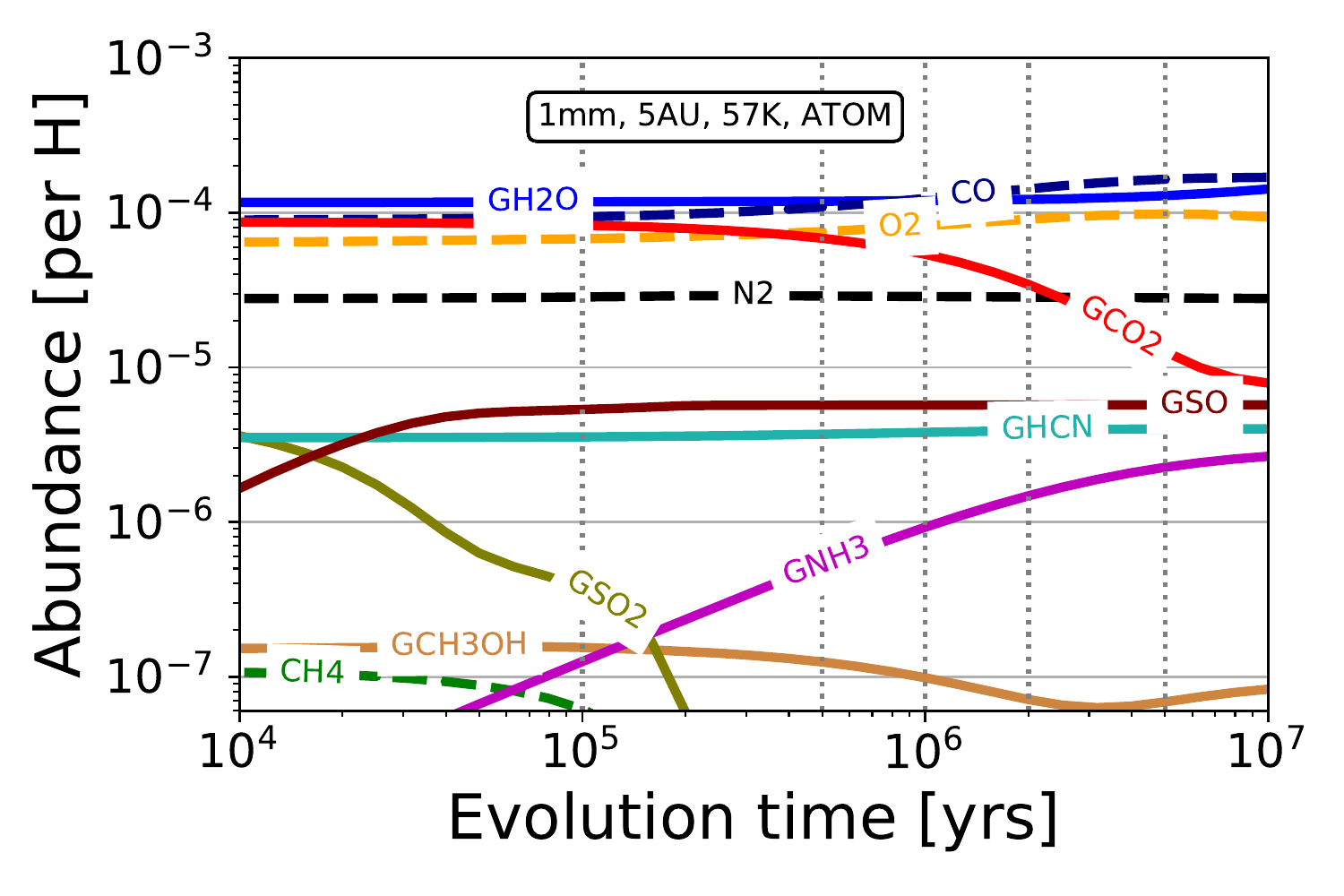}\\
    \includegraphics[width=.48\textwidth]{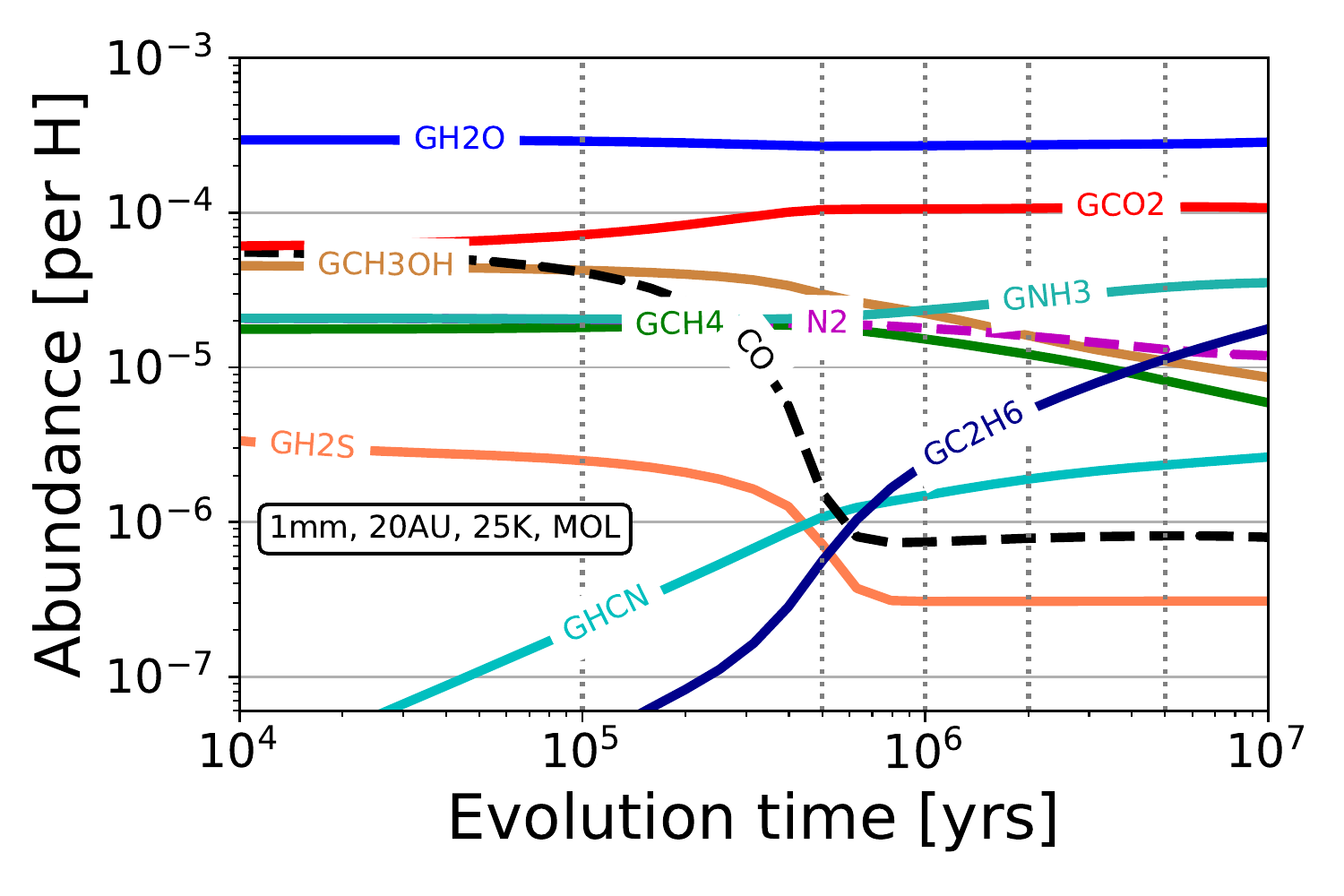}
    \includegraphics[width=.48\textwidth]{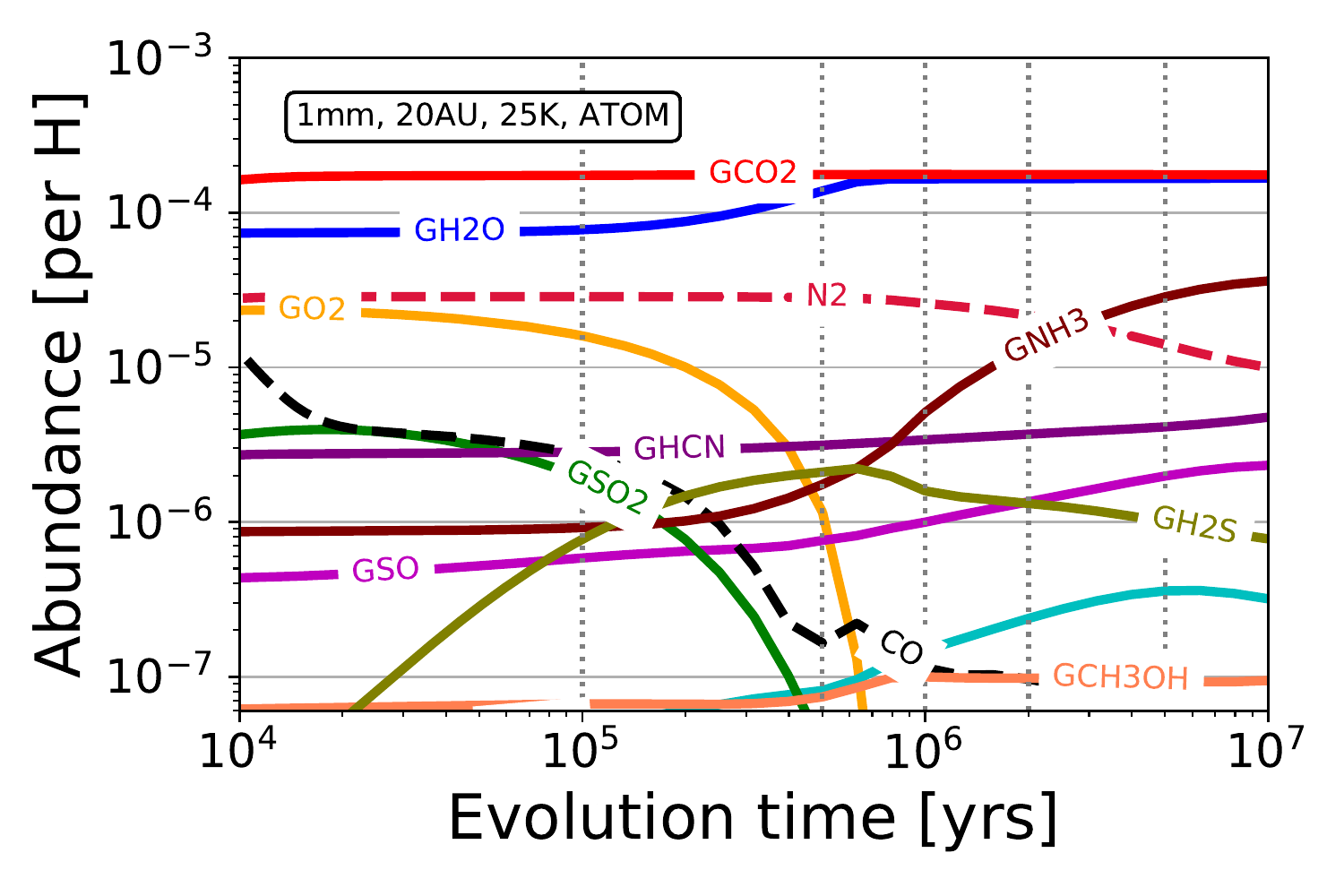}\\
    \includegraphics[width=.48\textwidth]{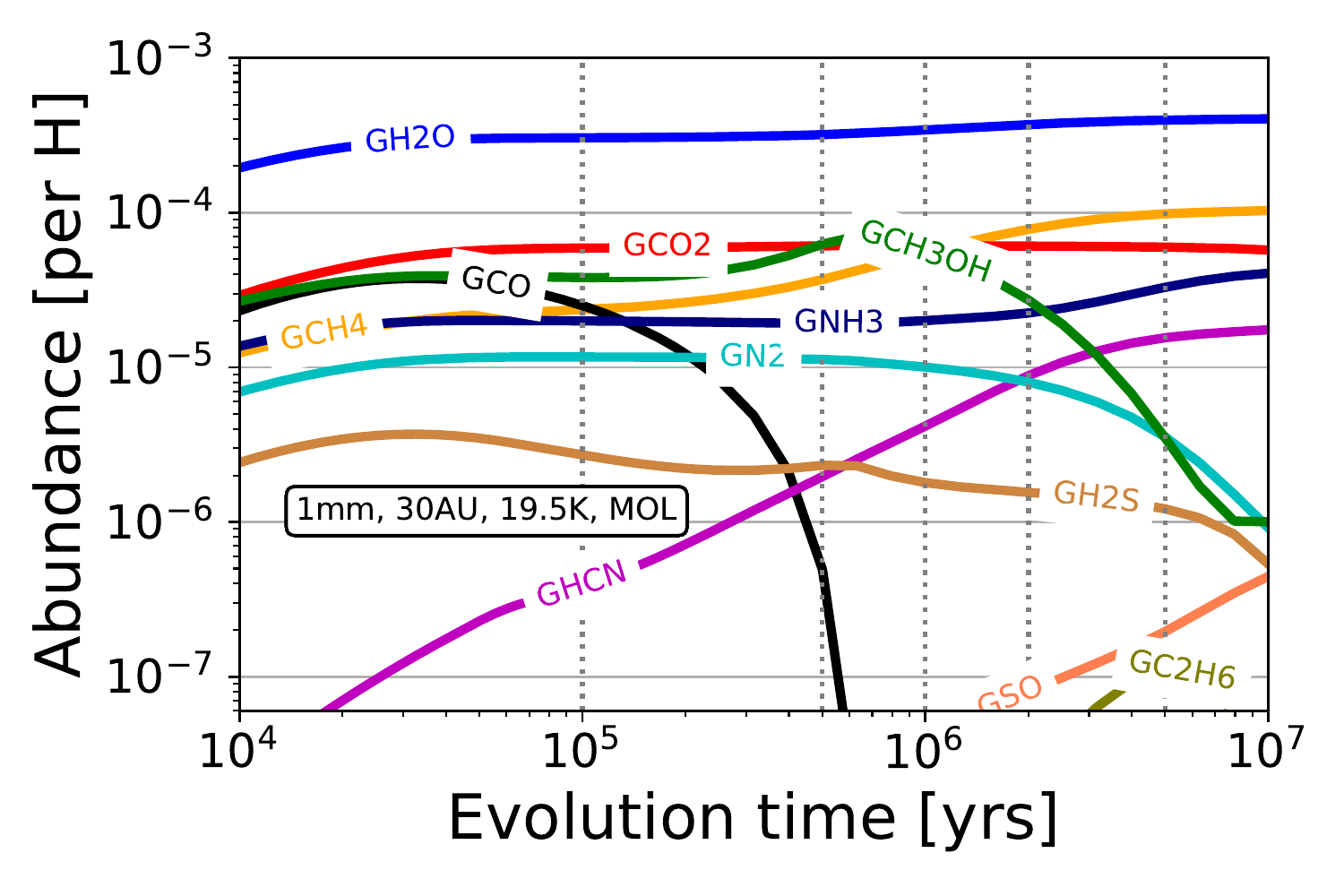}
    \includegraphics[width=.48\textwidth]{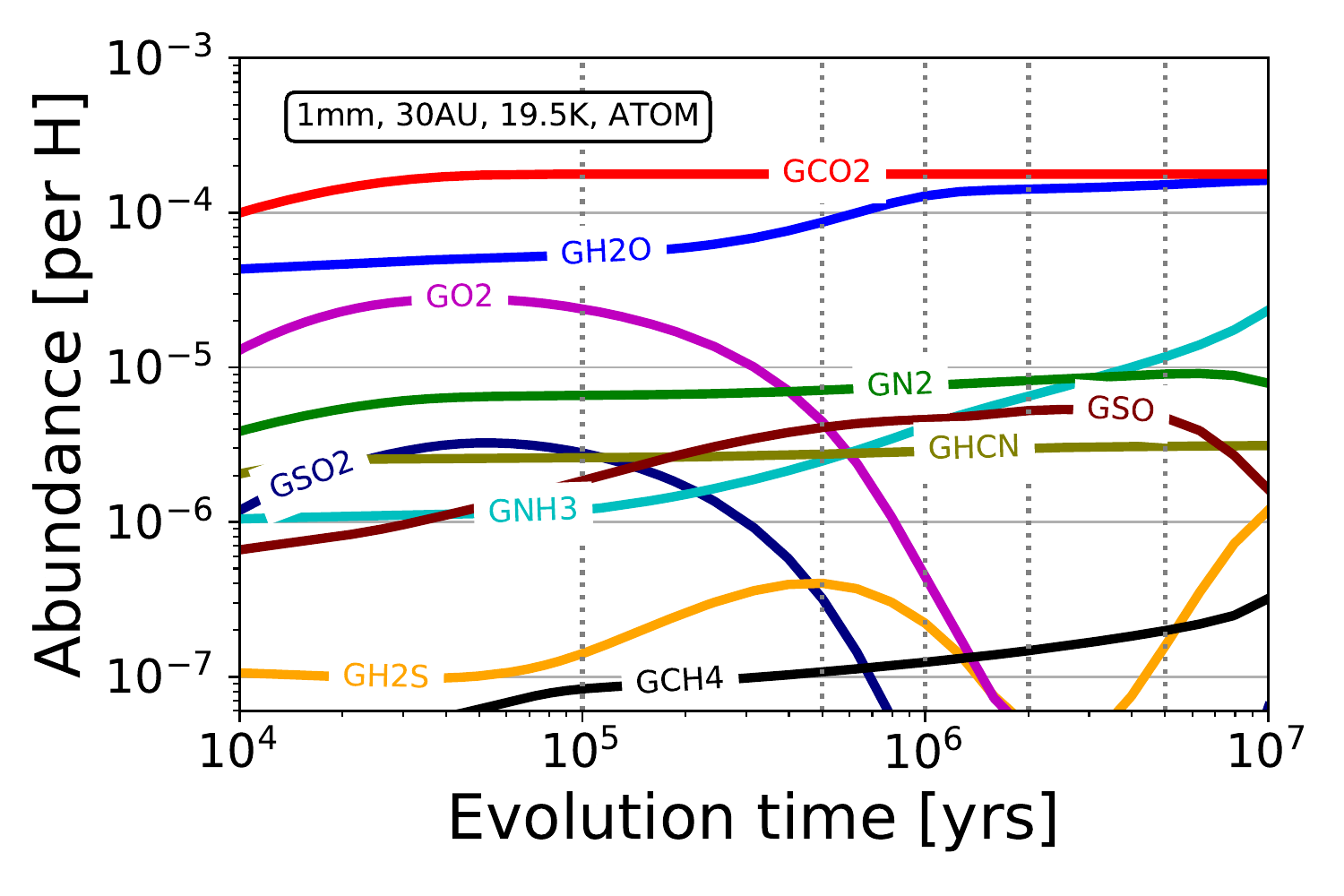}\\
    \caption{Evolution of chemical abundances at all radii (top-to-bottom: 1.5AU, 5AU, 20AU and 30AU) for a constant grain size of $R_\mathrm{grain}=1$mm. Solid profiles are for ices (species names starting with ``G''). Dashed profile are for gases. Left-hand panels: chemical models initiated with all elemental C, O, N and S in molecules (``MOL'' in insert). Right-hand panels: chemical models initiated with all elemental C, O, N and S as atoms (``ATOM'' in insert). Vertical dotted lines indicate evolution times used for analysis: 0.1Myr, 0.5Myr, 1Myr, 2Myr and 5Myr.}
    \label{1mm}
\end{figure*}

\end{appendix}



\end{document}